\documentclass[%
amsmath,
amssymb,
showkeys
]{revtex4-1}

\usepackage{graphicx}
\usepackage{dcolumn}
\usepackage{bm}

\usepackage{tcolorbox}

\usepackage{float}

\begin{document}

\preprint{APS/123-QED}

\title{Complex switching dynamics of interacting light in a ring resonator}


\author{Rodrigues D. Dikand{\'e} Bitha}
 \email{rodrigues.bitha@auckland.ac.nz}
\affiliation{Dodd-Walls Centre  for Photonic and Quantum Technologies, New Zealand}%
\affiliation{Department of Physics, University of Auckland, Private Bag 92019, Auckland, New Zealand}%
\affiliation{Department of Mathematics, University of Auckland, Private Bag 92019, Auckland, New Zealand}%
\author{Andrus Giraldo}%
\affiliation{School of Computational Sciences, Korea Institute for Advanced Study, Seoul 02455, Korea}%
\author{Neil G. R. Broderick}
\affiliation{Dodd-Walls Centre  for Photonic and Quantum Technologies, New Zealand}%
\affiliation{Department of Physics, University of Auckland, Private Bag 92019, Auckland, New Zealand}%
\author{Bernd Krauskopf}
\affiliation{Dodd-Walls Centre  for Photonic and Quantum Technologies, New Zealand}%
\affiliation{Department of Mathematics, University of Auckland, Private Bag 92019, Auckland, New Zealand}%

\date{\today}

\begin{abstract}
Microresonators are micron-scale optical systems that confine light using total internal reflection.  These optical systems have gained interest in the last two decades due to their compact sizes, unprecedented measurement capabilities, and widespread applications. The increasingly high finesse (or $Q$ factor) of such resonators means that nonlinear effects are unavoidable even for low power, making them attractive for nonlinear applications, including optical comb generation and second harmonic generation. In addition, light in these nonlinear resonators may exhibit chaotic behavior across wide parameter regions. Hence, it is necessary to understand how, where, and what types of such chaotic dynamics occur before they can be used in practical devices. We consider a pair of coupled differential equations that describes the interactions of two optical beams in a single-mode resonator with symmetric pumping. Recently, it was shown that this system exhibits a wide range of fascinating behaviors, including bistability, symmetry breaking, chaos, and self-switching oscillations. We employ here a dynamical system approach to identify, delimit, and explain the regions where such different behaviors can be observed. Specifically, we find that different kinds of self-switching oscillations are created via the collision of a pair of asymmetric periodic orbits or chaotic attractors at Shilnikov homoclinic bifurcations, which acts as a gluing bifurcation. We present a bifurcation diagram that shows how these global bifurcations are organized by a Belyakov transition point (where the stability of the homoclinic orbit changes). In this way, we map out distinct transitions to different chaotic switching behavior that should be expected from this optical device.
\end{abstract}

 \maketitle


\section{\label{intro} Introduction}

\begin{figure}[H]
\centering
\includegraphics[width=0.5\textwidth]{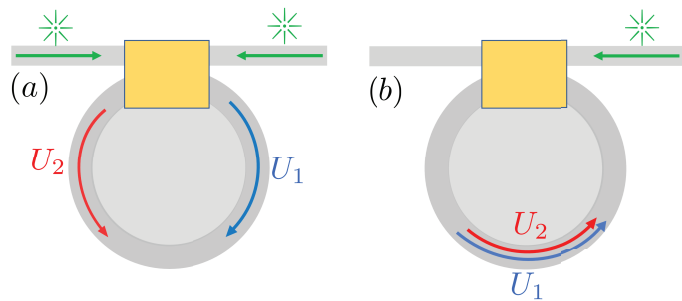}
\caption{\label{figure1} Schematics of different configurations of system~\eqref{eqt1}. (a) Dielectric micro-ring resonator: two identical beams are driven in opposite directions inside the resonator. (b) Long fiber ring resonator: the two orthogonally polarized modes propagate in the same direction inside the resonator.}
\end{figure}

Optical resonators are becoming increasingly important due to their nonlinear properties and potentially small sizes. While much work has been done on the propagation of light pulses in such cavities (motivated by the work of Leo \textit{et al.}~\cite{cavsol1} on temporal cavity solitons), we focus here on the dynamics of cavities where only a single longitudinal mode is excited. An optical resonator generally confines and stores light at specific frequencies, i.e., at the cavity's modes.  As resonators become smaller, the mode separation in frequency increases until only a single mode lies within the frequency range of interest, and they can be treated as quasi-single-moded with dispersive effects becoming negligible.  However, in general, two degenerate modes always exist, corresponding to either light counter-propagating in opposite directions in the resonator or co-propagating but with orthogonal polarizations. In either case, coupling between these two modes is practically unavoidable. Recently, Woodley \textit{et al.} have performed extensive experimental and theoretical studies of the counter-propagating case~\cite{delbino2017, woodley2018, selfswi} and have shown that the system exhibits a wide range of interesting behavior, including symmetry breaking, chaos, and self-switching oscillations, all resulting from the competition between self- and cross-modulation effects~\cite{kaplan} due to the Kerr nonlinearity~\cite{AGRAWAL1}.  Figure~\ref{figure1}(a) illustrates the principle of their experiments: two constant pump beams of equal powers are coupled into a single-mode ring resonator in opposite directions, and the output is observed as a function of time. Analytically, the propagation of light in such resonators is modeled by driven-dissipative wave equations with a linear loss and a Kerr nonlinearity given by~\cite{kaplan,woodley2018}:
\begin{equation}
\begin{aligned}
\frac{d U_1}{d t} &= [-1 + i (\alpha \vert U_1 \vert^2 + \beta \vert  U_2 \vert^2 - \Delta)]U_1 + f,
\\
\frac{d U_2}{d t} &= [-1 + i (\alpha \vert U_2 \vert^2 + \beta \vert U_1 \vert^2 - \Delta)] U_2 + f.
\label{eqt1}
\end{aligned}
\end{equation}
Here, $U_{1}$ and $U_{2}$ are the slowly varying complex envelopes of the two electric fields, $\alpha$ and $\beta$ are the self- and cross-phase modulation parameters~\cite{AGRAWAL1}, $f$ is the amplitude of the pump field inside the cavity, and $\Delta$ is the detuning between the frequency of the pump and the cavity mode. In system~\eqref{eqt1}, $U_{1}$, $U_{2}$ and $f$ have been normalized so that the Kerr nonlinearity is $1$ while the time is measured in units of decay time, meaning that the loss coefficient is $-1$. Note that exchanging $U_{1}$ and $U_{2}$ leaves system~\eqref{eqt1} invariant.

Notably, the system of equations~\eqref{eqt1} simultaneously describes the propagation in a ring resonator of two orthogonally polarised light fields with differing group velocities in the reference frame moving at their average speed. This configuration has been studied experimentally by Garbin {\it et al.}~\cite{bruno2020}; see Fig.~\ref{figure1}(b). The only difference in terms of the underlying model between their work and that of Woodley {\it et al.}~\cite{delbino2017, woodley2018, selfswi} lies in the different values for the self- and cross-phase modulation parameters in Eq.~\eqref{eqt1}.

The steady states of system~\eqref{eqt1} were first studied theoretically by Kaplan and Meystre~\cite{kaplan} in the context of nonlinear effects in Sagnac interferometers. In addition to optical bistability, they showed that this system might exhibit spontaneous symmetry breaking (where $U_1 \neq U_2$ in the steady states) and constructed a simplified phase diagram showing where this occurs. In 2017, Del Bino \textit{et al.}~\cite{delbino2017} experimentally observed both optical bistability and symmetry breaking in the context of two counter-propagating waves in a dielectric ring resonator~\cite{delbino2017}. They subsequently extended the mathematical analysis of the steady states~\cite{woodley2018,lewis2020} and showed, both numerically and experimentally, the existence of periodic solutions that undergo a period-doubling route to chaos.  In particular, they observed experimentally that the system could exhibit chaotic attractors with self-switching oscillations~\cite{selfswi}. In doing so, they considerably improved and extended the phase diagram of Kaplan and Meystre to include self-switching and chaotic regimes while still leaving some regions unexplored. 

In this paper, we provide a detailed theoretical analysis of the mechanism leading to the formation of chaotic dynamics and self-switching oscillations, and their organizing centers. To this end, we take a dynamical systems approach to study system~\eqref{eqt1}, focusing on its bifurcation structure. This allows
us to identify, delimit, and explain the different dynamical behaviors --- from simple to more complex --- that this system exhibits. We start our analysis where previous works~\cite{woodley2018,lewis2020,delbino2017,bruno2020} left off. Specifically, we focus on the organization of periodic orbits and their bifurcations in the two-dimensional parameter plane of intensity and detuning of the input field. To achieve this, we use numerical continuation techniques~\cite{kraus2007} implemented in AUTO-07p~\cite{auto} to complement our mathematical analysis. In doing so, we find that the relevant parameter region is organized by a Belyakov transition point~\cite{beltran1}, which leads to different types of nearby Shilnikov bifurcations~\cite{shill,shil2} with different symmetry properties due to the $\mathbb{Z}_2$-equivariance of system~(\ref{eqt1}). As we show, each of them is associated with a transition of a pair of periodic solutions and/or chaotic attractors via a symmetry-increasing gluing bifurcation that results in characteristic switching behavior. By continuing the different Shilnikov bifurcations in the parameter plane, we construct, for the first time to our knowledge, a bifurcation diagram representing the unfolding of the $\mathbb{Z}_2$-equivariant Belyakov transition. Since this type of codimension-two bifurcation is expected to occur in a wide class of nonlinear systems, these results should also be relevant to researchers in other fields. 

The paper is organized as follows. In Sec.~\ref{section01}, we discuss the symmetry properties of the system. Sec.~\ref{section1} then considers the bifurcations of the steady states; here, we derive the expressions for the emergence of optical bistability, symmetry breaking, and optical bistability of asymmetric steady states in the parameter plane. In Sec.~\ref{section2}, we study local and global bifurcations of both the steady states and periodic orbits and show how they are organized. Finally, in Sec.~\ref{belyakovsec}, we analyze the dynamics near the $\mathbb{Z}_2$-equivariant Belyakov transition and clarify the relevance of the associated Shilnikov bifurcations. Conclusions and an outlook are presented in Sec.~\ref{conclusion}.

\section{\label{section01} Rescaling and symmetry}

We rescale system~\eqref{eqt1} with the transformation $E_{1,2} =\sqrt{\alpha} U_{1,2}$ to introduce the ratio $B= \beta/\alpha$ of cross-phase and self-phase modulation as a single parameter; moreover, it is convenient for comparison with previous work \cite{kaplan,delbino2017,woodley2018,lewis2020,selfswi} to consider the rescaled input light intensity $F=f^2/\alpha$ as the main parameter in conjunction with the detuning $\Delta$. This yields the equivalent system 
\begin{equation}
\begin{aligned}
\frac{d E_1}{d t} &= [-1 + i ( \vert E_1 \vert^2 + B \vert E_2\vert^2 - \Delta)]E_1 + \sqrt{F},
\\
\frac{d E_2}{d t} &= [-1 + i ( \vert E_2 \vert^2 + B \vert E_1\vert^2 - \Delta)]E_2 + \sqrt{F},
\label{eq1}
\end{aligned}
\end{equation} 
for the rescaled electric fields $E_{1}, $ and $E_{2}$. 

Before turning to its bifurcation analysis, we first discuss some important properties of system~\eqref{eq1}. First of all, since $E_{1}, $ and $E_{2}$ are complex variables, the phase space is of dimension four. In particular, system~\eqref{eq1} can be written, in Cartesian form, as a vector field either for $(X_1,X_2,Y_1,Y_2)$ with $E_{1,2} = X_{1,2} + i Y_{1,2}$ or, in polar form, for $(r_1,r_2,\phi_1,\phi_2)$ with $E_{1,2} = r_{1,2} e^{i\phi_{1,2}}$. Note that $P_{1,2} = X_{1,2}^2 + Y_{1,2}^2 =  r_{1,2}^2$ is the intensity of the electric field $E_{1,2}$. Secondly, system~\eqref{eq1} is $\mathbb{Z}_2$-equivariant, that is, it remains unchanged under the transformation of exchanging the two electric fields
$$\eta: (E_1,E_2) \mapsto (E_2,E_1).$$ 
Hence, a solution of system~\eqref{eq1} is either mapped to itself (as a set) by $\eta$, in which case we refer to it as a \emph{symmetric} solution, or it is an \emph{asymmetric} solution; importantly, \emph{asymmetric} solutions come in pairs that are each other's images under the symmetry transformation $\eta$. An important object is the \emph{fixed-point subspace} of an equivariant systems~\cite{equivariant,golubitsky,andrus2021}, which for system~\eqref{eq1} is 
$$
\begin{array}{cc}
    \text{Fix}_\eta := \{(E_1,E_2) \in \mathbb{C} \times \mathbb{C} | \; E_1 = E_2 \}.
\end{array}
$$
The fixed-point subspace $\text{Fix}_\eta$ is two-dimensional, invariant under the flow, and consists of all solutions that are fixed pointwise under $\eta$. Trajectories in $\text{Fix}_\eta$ are symmetric and characterized by having the same intensities $P_{1}(t) = P_{2}(t)$ for all time. The dynamics of system~\eqref{eq1} in $\text{Fix}_\eta$ reduces to the single equation
\begin{equation}
\frac{d E}{d t} = [-1 + i (B + 1)\vert E \vert^2 - i \Delta]E + \sqrt{F},
\label{eqt101}
\end{equation} 
where $E=E_1=E_2$. Note that any steady state of system~\eqref{eq1} that is symmetric must lie in $\text{Fix}_\eta$; otherwise it is asymmetric. To determine the stability of steady states, we use the Jacobian matrices $J$ of system~\eqref{eq1} and $J_S$ of system~\eqref{eqt101} in their Cartesian form, as given in Appendix~\ref{appendix1}. 

For a symmetric periodic solution $\Gamma$ with (minimal) period $T$ there are actually two possibilities of how $\eta$ maps it as a set:
\begin{itemize}
    \item $\Gamma$ is a \emph{fixed-point periodic solution} if $\Gamma \subset \text{Fix}_\eta$, which means that all points $\Gamma(t)$ are invariant under $\eta$; this case does not actually occur in system~\eqref{eq1} (see Appendix~\ref{appendix1});
    \item otherwise, $\Gamma$ is an \emph{$\mathbf{S}$-invariant periodic solution}; in this case, $\emptyset \neq \Gamma \cap \text{Fix}_\eta \neq \Gamma$ and $\Gamma(t)$ is mapped to itself by $\eta$ only in combination with a shift in the time $t$.
\end{itemize}
We remark that periodic orbits are global objects that, in general, need to be found numerically; we use the continuation package AUTO-07p~\cite{auto} to find periodic orbits, determine their stability, and detect and continue their (global) bifurcations.

\section{\label{section1} Steady-state bifurcations}

Previous works~\cite{delbino2017,woodley2018,bruno2020} have shown that system~\eqref{eq1} exhibits spontaneous symmetry breaking and bistability of symmetric and asymmetric steady states. We now analytically derive the curves of local bifurcations underlying these phenomena. Unlike in previous work~\cite{woodley2018,lewis2020}, our formulas for the loci of the transition to bistability and of symmetry breaking are presented here in terms of all parameters, including the ratio $B$ of cross- and self-phase modulation and the phase $\phi$. Moreover, we also provide an expression for the locus of the boundary of the region with multiple stable asymmetric steady states.

\subsection{\label{subsection1} Optical bistability: saddle-node bifurcations of symmetric steady states}

Optical bistability is a fundamental concept in optics, where an optical system exhibits two stable states for the same values of the parameters~\cite{Optical_Bistability_Gibbs}. For systems with optical Kerr nonlinearity, this phenomenon is underpinned by two successive saddle-node bifurcations involving two different stable solutions of the system. We first consider the existence of two stable steady states in the fixed-point subspace $\text{Fix}_\eta$; that is, we first focus on the bistability of symmetric steady states. To find the locus of saddle-node bifurcations in $\text{Fix}_\eta$, we perform the analysis in polar coordinates, with $E_{1}=E_{2}=r e^{i\phi}$. Then system~\eqref{eqt101} takes the form
\begin{equation}
  \label{eq2}
  \begin{aligned}
    \frac{d r}{d t} &= \sqrt{F} \cos{\phi} - r,\\        
    \frac{d \phi}{d t} &=- \frac{\sqrt{F}\sin{\phi}}{r} - \left(B+1 \right)r^2 + \Delta.
  \end{aligned}
\end{equation}
It follows directly from system~\eqref{eq2} that any symmetric steady state satisfies $r=\sqrt{F}\cos{\phi}$. A necessary condition for a saddle-node bifurcation to occur is that the Jacobian matrix $J_S$ of system~\eqref{eq2}, evaluated at a steady state, has a zero eigenvalue, which means $\det(J_S)=0$. In polar coordinates
\begin{equation} 
\det (J_S) = 1 + \tan^2{\phi} -(B+1)F\sin{2\phi},
\label{eq3}
\end{equation}
where the condition $r=\sqrt{F} \cos{\phi}$ for a symmetric steady states has been used. 

\paragraph*{Parameterization of the saddle-node bifurcation $\mathbf{S}$.}
Solving for the zeros of system~\eqref{eq2} and Eq.~\eqref{eq3} yields the parametrization of the input power $F_S$ and the detuning $\Delta_S$,

\begin{equation}
\textbf{S}: \left( F_S(\phi),  \Delta_S(\phi)  \right) = \left(\frac{1+\tan^2{\phi}}{(B+1)\sin{2\phi}},\quad  \frac{1+2\sin^2{\phi}}{\sin{2\phi}} \right),
  \label{eq4}
\end{equation}
with $\phi \in (0, \pi/2) $. Equation~\eqref{eq4} defines the locus of the saddle-node bifurcations $\mathbf{S}$ of steady states in $\text{Fix}_\eta$, which delimits the region of optical bistability of symmetric states in the $(F,\Delta)$-plane. 
The locus $\mathbf{S}$ has a minimum at $\phi=\pi/6$, at which point there is a cusp bifurcation in the 
$(F, \Delta)$- plane at
\begin{equation}
    \textbf{CP}_S: \left( F_{CP_S}\left(\dfrac{\pi}{6}\right),  \Delta_{CP_S}\left(\dfrac{\pi}{6}\right) \right)=  \left(\dfrac{8}{3\sqrt{3}(B+1)}, \sqrt{3} \right).
    \label{cusps}
\end{equation}
Hence, a necessary condition for bistability is that $F>F_{CP_S}$ and $\Delta>\Delta_{CP_S}$. We note that $\Delta_S$ is independent of $B$ implying that there is a minimum detuning needed to observe bistability that is independent of the relative strengths of the self- and cross-phase modulation terms.

\subsection{\label{subsection2} Spontaneous symmetry breaking: pitchfork bifurcation of symmetric steady states}

Spontaneous symmetry breaking for steady states of system~\eqref{eq1} was first predicted  by Kaplan and Meystre~\cite{kaplan}. Above a certain threshold of the input power and/or detuning, the symmetric states lose their stability in favor of two asymmetric states through a pitchfork bifurcation $\mathbf{P}$. The newly created asymmetric steady states emerge transversely to the subspace $\text{Fix}_\eta$. 


We can determine the boundaries of the symmetry-breaking region by considering the full Jacobian $J$ (see Appendix~\ref{appendix1}) of system~\ref{eq1} at the symmetric steady states~\cite{andrus2020,andrus2021}.
The condition for $\mathbf{P}$ is that one of the eigenvalues of the $J$ becomes zero, that is, $\det(J)=0$. In polar coordinates, the determinant of $J$ evaluated at any symmetric state takes the form
\begin{equation}
    \det(J)= \det (J_S)(1 + \tan^2{\phi} +(B-1)F\sin{2\phi}).
    \label{jacobian}
\end{equation}
The first factor, $\det(J_S)$, contains the information on the bistability region as discussed in Section~\ref{subsection1}. The second factor gives us insight into other bifurcations whose \textit{center manifold}~\cite{Kuznesovbook} is not contained in $\text{Fix}_\eta$. One readily sees from the second factor of det($J$) that higher bifurcation degeneracies occur when:
\begin{itemize}
    \item $B=0$: the two factors of $\det(J)$ are equal, and the system only exhibits symmetric steady states. In this case, system~\eqref{eqt1} models the dynamics of two uncoupled electric fields. 
    \item  $B=1$: the second factor is strictly positive, which implies that it does not generate bifurcations. 
\end{itemize}

In these two cases, the only local bifurcations of symmetric steady states in system~\eqref{eq1} are the saddle-node bifurcations we studied in section~\ref{subsection1}.

\paragraph*{Parameterization of the pitchfork bifurcation $\mathbf{P}$.}
When $B>0$ and $B\neq 1$, system~\eqref{eq1} exhibits pitchfork bifurcations for suitable values of the parameters. In this case, the necessary condition arises from the second factor of $\text{det}(J)$, that is,
\begin{equation}
1 + \tan^2{\phi} +(B-1)F\sin{2\phi}=0.
\label{eq6}
\end{equation}

By solving for the zeros Eqs.~\eqref{eq2} and~\eqref{eq6}, we obtain the locus of the pitchfork bifurcations delimiting the region of symmetry breaking as:
\begin{equation}
\textbf{P}: \left(  F_P(\phi),  \Delta_P(\phi)  \right)=  \left(  \frac{1+\tan^2{\phi}}{(1-B)\sin{2\phi}},\quad  \frac{\frac{B+1}{1-B}+2\sin^2{\phi}}{\sin{2\phi}} \right).
\label{eq7}
\end{equation}
The fact that $F_P$ must be positive forces the domain of $\phi$ to be $\phi \in (0, \pi/2)$, when $0<B<1$, and $\phi \in (-\pi/2, 0) $, when $B>1$. 

Notice that 
$$\lim_{B \to 1^\pm} F_P = + \infty,\quad  \phi \neq 0,\, \pm \pi/2 $$
%
In other words, the pitchfork bifurcation appears under any small perturbation of $B$ from this limiting case, $B=1$. 

The values $F_P$ and $\Delta_P$ reach their respective minima for different values of the phase at
$$\phi_{SB}=\frac{\pi}{6} \quad \text{and} \quad \phi_{SB}^*=\arctan{\left(\sqrt{\frac{B+1}{3-B}}\right)},$$
respectively. These values define the minimum thresholds for symmetry breaking when sweeping the cavity with either the detuning or input power; more precisely,
\begin{itemize}
    \item  The minimum input power necessary to achieve symmetry breaking is $$F_P\left(\phi_{SB}\right)=\dfrac{8}{3\sqrt{3}(B-1)}.$$
    For the particular case $B=2$, the threshold of symmetry breaking is $F_{\text{SB}}=\frac{8}{3\sqrt{3}}$, as demonstrated in Ref.~\cite{woodley2018}.
    \item  On the other hand, a necessary condition on the detuning for symmetry breaking to occur is that the detuning must be larger than $$\Delta_P\left(\phi_{SB}^*\right)=\dfrac{\sqrt{(1 + B)(3 - B)}}{B-1}.$$ 
    Notice that this condition only holds when $B<3$. As $B \rightarrow 3$, the phase $\phi_{SB}^*$ goes to infinity, and the minimum detuning for symmetry breaking $\Delta_P(\phi_{SB}^*)$ tends to zero. Thus when $B>3$, spontaneous symmetry breaking can be observed for any negative detuning for suitable pump power. 
\end{itemize}

\begin{figure}
\includegraphics{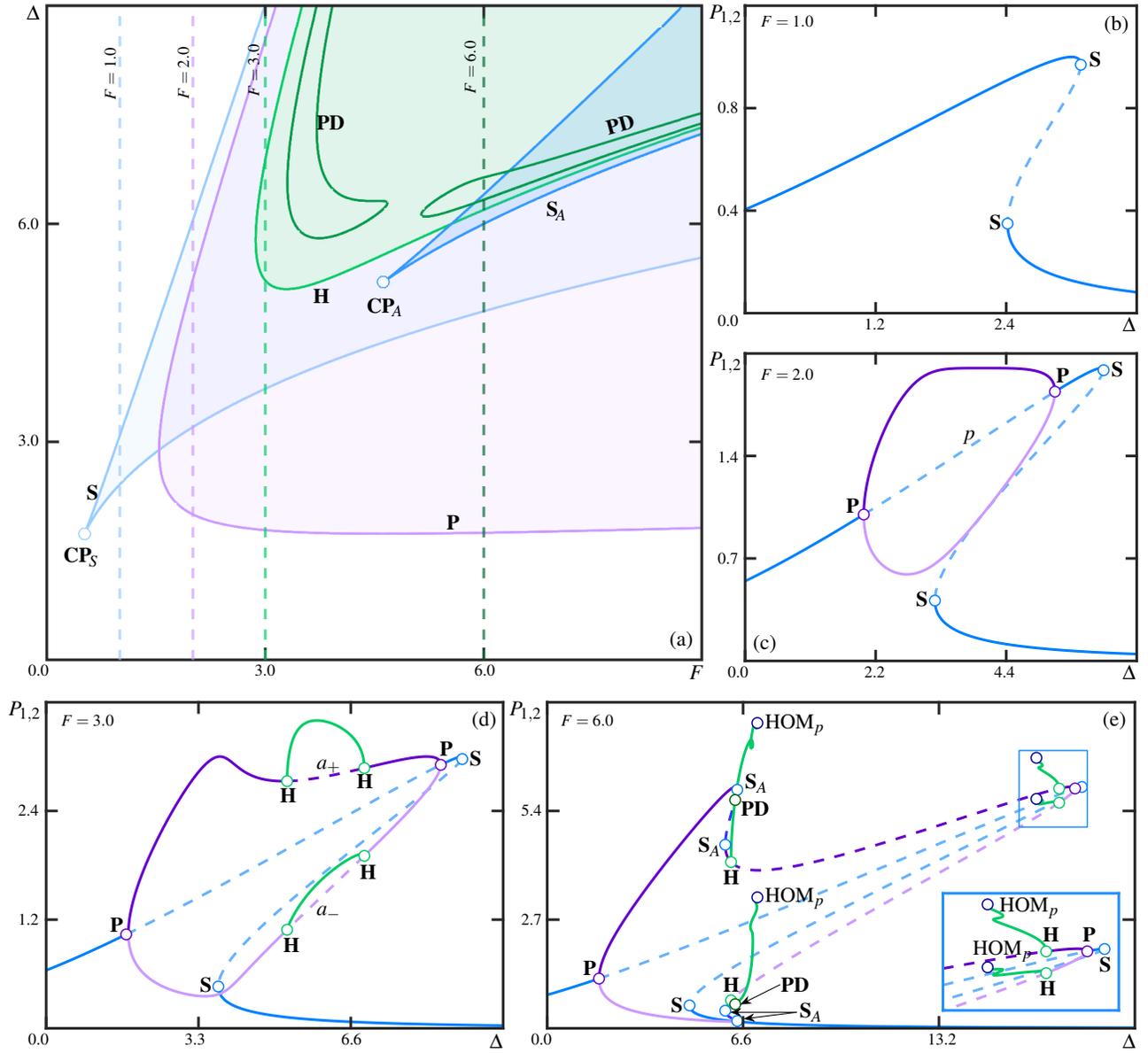}
\caption{\label{fig2} (a) Two-parameter bifurcation diagram of system~\eqref{eq1} in the $(F, \Delta)$-plane. Shown are the curves: $\mathbf{S}$ (light blue) and $\mathbf{S}_A$ (blue) of saddle-node bifurcation of the symmetric and asymmetric steady states, respectively, and of pitchfork $\mathbf{P}$ (purple), Hopf $\mathbf{H}$ (green) and period-doubling $\mathbf{PD}$ (dark green) bifurcations. Regions of different dynamical behaviors are highlighted in color. (b)--(e) one-parameter bifurcation diagrams of system~\eqref{eq1} in $\Delta$ for fixed values of $F$ as shown, with symmetric (blue) and asymmetric steady states (dark and light purple), and periodic orbits (green) represented by their maxima. Stable and unstable solutions are represented by solid and dashed curves, respectively.}
\end{figure}

\subsection{\label{subsection3} Optical bistability: saddle-node bifurcations of asymmetric steady states} 

To study the dynamics of asymmetric steady states of system~\eqref{eq1}, we investigate the corresponding bifurcation conditions analytically. Steady states of system~\eqref{eq1} obey the two Lorentzian equations (see Appendix \ref{appendix2})

\begin{subequations}
\begin{align}
 F = P_{1} + (P_{1} + B P_{2} -\Delta)^2P_{1},
 \label{eq:11a}
\\
 F = P_{2} + (P_{2} + B P_{1} -\Delta)^2P_{2},
\label{eq:11b}
\end{align}
\label{eq:11}
\end{subequations}
where $P_{1,2}=|E_{1,2}|^2$ are the output intensities. Rearranging Eqs.~\eqref{eq:11} shows that the output intensities $P_{1,2}$, are solutions of the equations:
\begin{subequations}
\begin{align}
0 &=  (P_1 - P_2)[(P_1 + P_2 - \Delta)^2 + 1 - (B-1)^2 P_1 P_2],
\label{eq:12a}\\
F & =\frac{[(B+1)(P_1 + P_2) -2\Delta][(P_1 + P_2 - \Delta)^2 + 1]}{B-1}.
\label{eq:12b}
\end{align}
\label{eq:12}
\end{subequations}  
Since we are only interested in asymmetric steady states, the first factor of Eq.~\eqref{eq:12a}, which determines the condition for symmetric steady states $P_1=P_2$, can be ignored. With the transformation $S=P_1 + P_2$, the system of coupled equations~\eqref{eq:12} can be reduced to 
\begin{equation}
 F(B - 1) = [(B+1)S -2\Delta][(S - \Delta)^2 + 1].
\label{eq:13}
\end{equation} 
Notice that the right-hand side of Eq.~\eqref{eq:13} is a third-order polynomial in $S$, whose zeros depend on $\Delta$ and $B$. Therefore, one can determine two threshold values, $S_-$ and $S_+$, which delimit a region with three possible real solutions of $S$. This region is well-known in optics as a hysteresis cycle of systems with cubic nonlinearity~\cite{Optical_Bistability_Gibbs}. One determines the threshold values $S_{\pm}$ by solving for the zeros of the partial derivative $\partial F/ \partial S $ to obtain
\begin{equation}
    S_{\pm}=\dfrac{2(B+2)\Delta \pm \sqrt{(B-1)^2 \Delta^2 - 3(B+1)^2}}{3(B+1)}.
    \label{eq:16}
\end{equation}
Hence, for $\Delta> \sqrt{3}(B+1)/(B-1)$ and $B\neq 1$, two asymmetric steady states of $P_{1,2}$ collide at $S_-$ and $S_+$. By replacing $S_{\pm}$ in Eq.~\eqref{eq:13}, we obtain the locus
\begin{equation}
  \begin{aligned}
\textbf{S}\!_A: F_{S_A}(\Delta)=\dfrac{2(B-1)}{27(B+1)^2}\Big\{ \left[(B-1)^2\Delta^2 + 9(B+1)^2\right]\Delta & \\        
    \pm \left[\Delta^2 - 3(B+1)^2/(B-1)^2\right]^{3/2}  \Big\} &, 
    \label{eq:17}
  \end{aligned}
\end{equation}
of saddle-node bifurcation of asymmetric steady states in the ($F,\Delta$)-plane.

From Eq.~\eqref{eq:17}, the threshold value of the bistability region of asymmetric steady states in the ($F,\Delta$)-plane is determined; it is given by the cusp bifurcation of the asymmetric steady states
\begin{equation}
    \textbf{CP}\!_A: \left( F_{CP_A}, \Delta_{CP_A} \right)= \left( \dfrac{8}{3\sqrt{3}}(B+1), \quad \dfrac{B+1}{B-1}\sqrt{3} \right).
    \label{cpa}
\end{equation}

As we show in the following section, the region enclosed by the saddle-node bifurcation of asymmetric steady states overlaps with other regions; this creates regions of considerable multi-stability in the $(F,\Delta)$-plane.

\section{ \label{section2} Bifurcation diagram of steady states and periodic solutions}
 We now use the information from the previous section in conjunction with numerical continuation to find local and global bifurcations of steady states and periodic orbits with the software package AUTO-07p~\cite{auto}. To compare our results with previous studies~\cite{woodley2018,delbino2017,lewis2020}, we keep the ratio between the cross- and self-phase modulations constant at $B=2$.

Figure~\ref{fig2} shows the bifurcation diagram of local bifurcations of steady states and periodic orbits of system~\eqref{eq1} in the $(F,\Delta)$-plane (a), as well as one-parameter bifurcation diagrams (b)--(e) in the detuning $\Delta$ for fixed values of the pump power $F$. In panels~(b)--(e), stable and unstable symmetric solutions are represented by solid and dashed blue curves, respectively. We note that all the unstable steady states of system~\eqref{eq1} are saddle-foci. Asymmetric steady states are shown in purple, and the green curves represent the maximum values of periodic solutions. Notice that each bifurcation diagram in panels (b)--(e) illustrates a vertical cut of the $(F,\Delta)$-plane in panel (a), as indicated by the dashed lines.

\subsection{ \label{twoparam}Two-parameter bifurcation diagram}

Figure~\ref{fig2}(a) shows the $(F,\Delta)$-plane of system~\eqref{eq1} with regions of simple dynamics in different colors, bounded by different bifurcation curves. In the white region, the system exhibits a single symmetric steady state with $P_1=P_2$. The bistability and symmetry-breaking regimes of symmetric steady states are shown in light blue and purple, respectively, and the region of bistability of asymmetric steady states is shown in blue. Note that these regions are bounded by the bifurcation curves $\mathbf{S}$, $\mathbf{P}$ and $\mathbf{S}\!_A$, respectively, as derived in Section~\ref{section1}. Notice also the cusp bifurcation points $\mathbf{CP}\!_S$ and $\mathbf{CP}\!_A$, which represent the thresholds of the bistability regimes of symmetric and asymmetric steady states, respectively, as obtained in Eqs.~\eqref{cusps} and~\eqref{cpa}.
 
In addition to these local bifurcations derived analytically, we numerically find that the asymmetric steady states exhibit Hopf bifurcations $\textbf{H}$, leading to a region with periodic orbits. The curve $\textbf{H}$ in  Fig.~\ref{fig2}(a) marks the onset of oscillations, and it bounds the green region. Inside this region, we find that periodic orbits emerging from $\textbf{H}$ exhibit further bifurcations, including cascades of period-doubling bifurcations. Figure~\ref{fig2}(a) already shows the loci $\textbf{PD}$ of the first two period-doubling bifurcations. Different bifurcations that occur in the region delimited by the curve $\textbf{H}$ are introduced and discussed in more detail in Section~\ref{subsection5}.

\subsection{\label{dlow} One-parameter bifurcation diagrams}
Notice that all the regions with different dynamics in Fig.~\ref{fig2}(a) overlap to form regions of different types of multi-stability. To understand the relevance of these different regions, we show, in Figs.~\ref{fig2}(b)--(e), for several values of $F$, one-parameter bifurcation diagrams of $P_{1,2}$ as a function of  $\Delta$. The values of $F$ have been carefully chosen to showcase the gradual increase in complexity of the dynamics of the output intensities $P_{1,2}$ with increments of $F$. 

For $F=1.0$ as in Fig.~\ref{fig2}(b), system~\eqref{eq1} undergoes two saddle-node bifurcations $\textbf{S}$. Notice that there is only one branch because the system is in the symmetric regime with $P_1=P_2$. The two bifurcations $\textbf{S}$ lead to a $\Delta$-range where two stables symmetric steady states (with different output intensities) and a saddle symmetric steady state exist simultaneously; see Fig.~\ref{fig2}(b). More specifically, the three-dimensional \textit{stable manifold} of this saddle steady state, which is the set of all points in phase space that converge toward this state in forward time, separates the basins of attraction of these steady states. Initial conditions on one side of this manifold converge to the stable steady state with a high output intensity while, on the other side, they converge to the stable steady state with a low output. This phenomenon, well-known as optical bistability, was observed experimentally~\cite{delbino2017}. The region where these two stable symmetric steady states coexist is delimited by the curve $\textbf{S}$ and shaded in light blue in Fig.~\ref{fig2}(a).

For $F=2.0$ as in Fig.~\ref{fig2}(c), the symmetric steady state with the high output intensity becomes unstable at a supercritical pitchfork bifurcation  $\textbf{P}$. Hence, two stable (and asymmetric) steady states with $P_1 \neq P_2$ are created and represented by two branches. This phenomenon is the well-known spontaneous symmetry breaking~\cite{delbino2017,bruno2020}. Eventually, as $\Delta$ increases, the higher symmetric steady state regains its stability at a second supercritical pitchfork bifurcation $\textbf{P}$, where the two asymmetric steady states disappear. The region of symmetry-broken steady states is delimited by the curve $\textbf{P}$ in the ($F$, $\Delta$)-plane in Fig.~\ref{fig2}(a).

\begin{figure}
\centering
\includegraphics[]{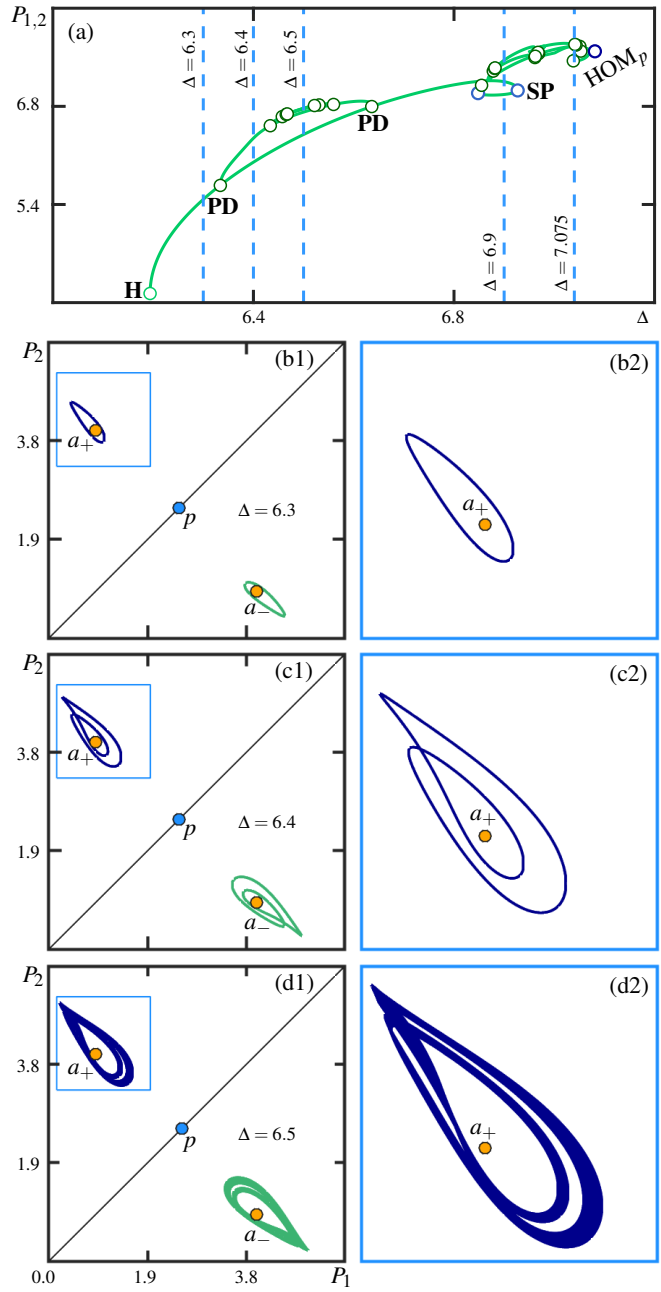}
\caption{\label{fig3} (a) One-parameter bifurcation diagram (of a pair) of asymmetric periodic orbits in $\Delta$ for $F=6.0$. (b)--(d) Associated pairs of attractors  in the ($P_1$,$P_2$)-plane for $\Delta=6.3$, $\Delta=6.4$ and $\Delta=6.5$, respectively. Symmetric (blue) and asymmetric (gray) equilibria are indicated as points, and framed regions of panels (b1)--(d1) are enlarged in~(b2)--(d2). }
\end{figure}
\begin{figure}
\centering
\includegraphics[]{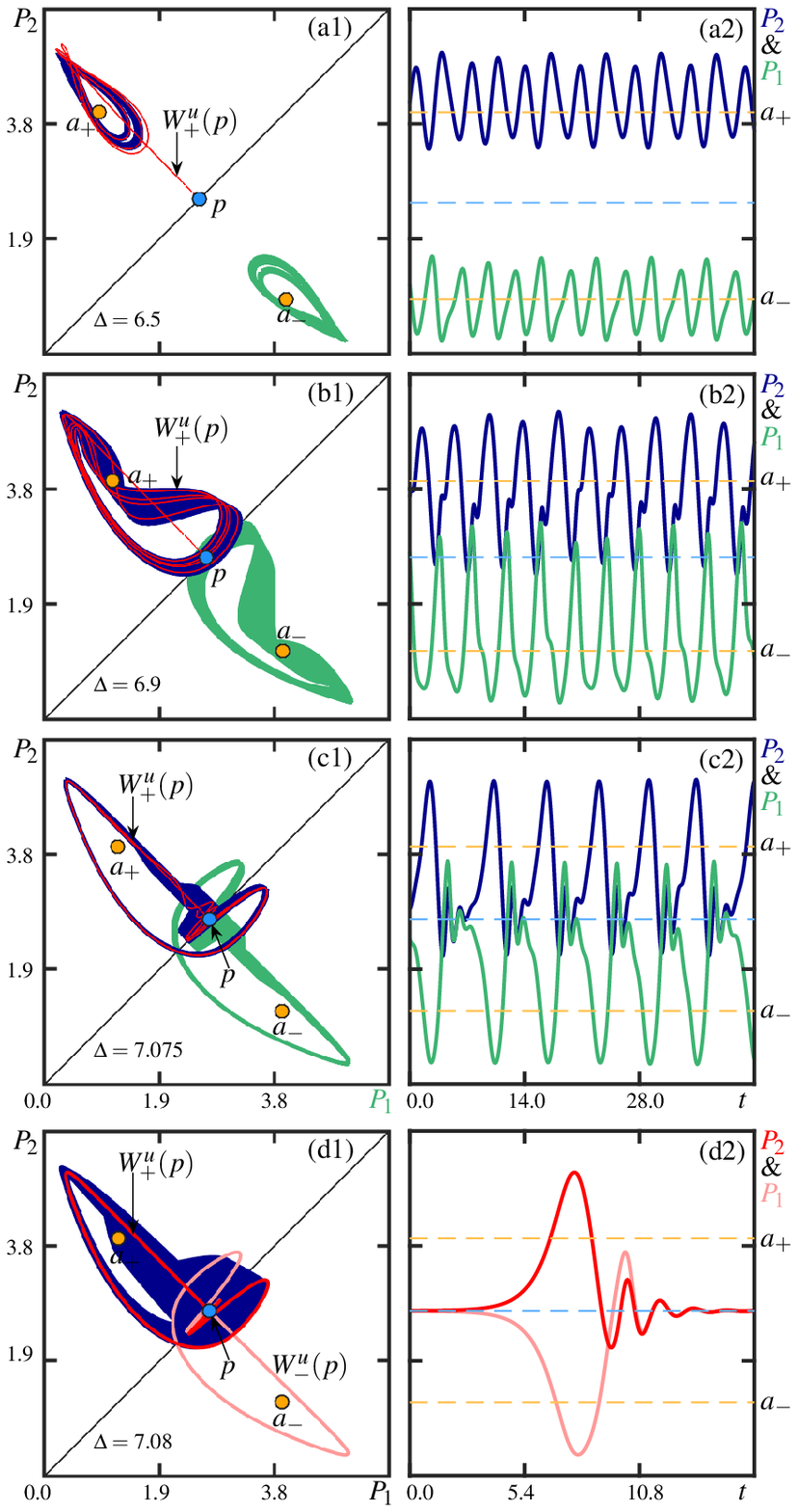}
\caption{\label{fig4} (a1)--(c1) Pairs of asymmetric chaotic attractors and (a2)--(c2) their associated temporal traces for $F=6.0$ and $\Delta=6.5$, $\Delta=6.9$, $\Delta=7.075$, respectively. (d1) Pair of Shilnikov homoclinic orbits and (d2) corresponding temporal profiles for $F=6.0$ and $\Delta=7.08$. Also shown in (a1)--(c1) is $W_+^u(p)$.}
\end{figure}

Notice that the two regions delimited by the curves $\textbf{S}$ and $\textbf{P}$ overlap in Figs.~\ref{fig2}(a); hence, there is a range of multi-stability in panel (c) where the lower symmetric steady state and the two asymmetric steady states exist and are stable simultaneously. As a result, when scanning system~\eqref{eq1} for $F=2.0$, from large to small values of $\Delta$, the output intensities $P_{1}$ and $P_{2}$ both switch from the lower symmetric steady state to one of the two different asymmetric steady states. By carefully choosing the initial condition, one determines in numerical simulations whether $P_1$ or $P_2$ switches to the asymmetric steady states with the higher output intensity. However, in an experiment, the dominant asymmetric state is often determined by small asymmetries of the system~\cite{bruno2020}.

For $F=3.0$ as in Fig.~\ref{fig2}(d), the one-parameter bifurcation diagram reveals that each asymmetric steady state undergoes a pair of supercritical Hopf bifurcations $\textbf{H}$. After the first Hopf bifurcation $\textbf{H}$, a pair of stable and asymmetric periodic orbits emerges. As $\Delta$ increases, this pair of periodic orbits disappears at the second Hopf bifurcation $\textbf{H}$. In Fig.~\ref{fig2}(a), there is an area where the bistability regime of symmetric states overlaps with the green region of periodic orbits bounded by the curve $\textbf{H}$ of Hopf bifurcation. In Fig.~\ref{fig2}(d), this area corresponds to a parameter range of multi-stability, where stable periodic orbits coexist with the stable and symmetric steady state of low output intensity.

For $F= 6.0$ as in Fig.~\ref{fig2}(e), each of the asymmetric steady states undergoes two saddle-node bifurcations, leading to a hysteresis loop delimited by the points $\textbf{S}_A$. In this $\Delta$-range, system~\eqref{eq1} exhibits two pairs of stable and asymmetric steady states that coexist. Hence, the system becomes multi-stable past the point $\textbf{S}_A$ for increasing $\Delta$, with coexistent symmetric and asymmetric steady states and periodic solutions past $\textbf{H}$. Moreover, the pair of periodic orbits that emerge at the first point $\textbf{H}$ in Fig.~\ref{fig2}(e) undergo multiple sequences of period-doubling $\textbf{PD}$ and saddle-node $\textbf{SP}$ bifurcations, and the branch of periodic orbits ends at a pair of Shilnikov homoclinic bifurcations~\cite{shill, shil2} denoted $\text{HOM}_p$. The same happens for decreasing $\Delta$ with the branches of periodic orbits emerging at the second pair of Hopf bifurcations. Interestingly, we find a region between the points $\text{HOM}_p$ in Fig.~\ref{fig2}(e), where the low symmetric steady state coexists with chaotic attractors of different symmetry properties. This agrees with the one-parameter bifurcation diagram presented in Ref.~\cite[Fig.~2]{selfswi}, where chaotic and complex dynamics have been observed in numerical simulations. In the remainder of the paper, we consider in detail the emergence of these chaotic attractors and their symmetry properties.

\begin{figure}
\centering
\includegraphics{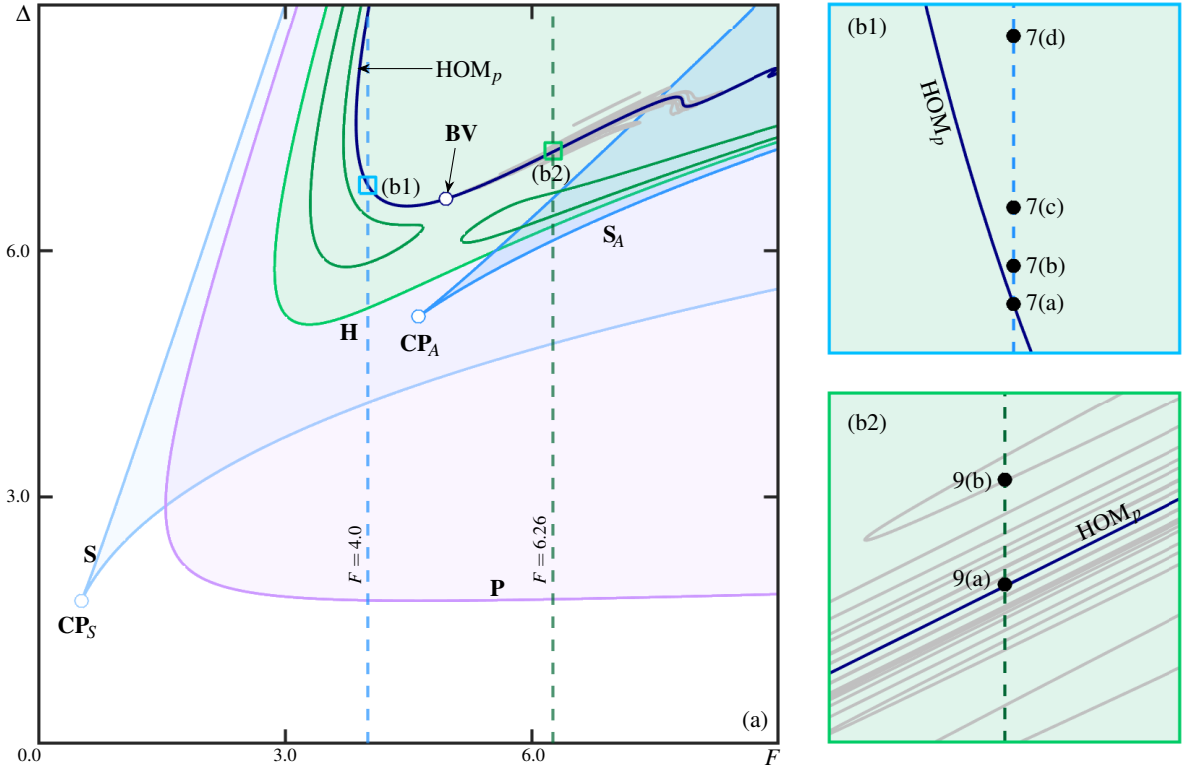}
\caption{\label{fig5} (a) Two-parameter bifurcation diagram of system~\eqref{eq1} in the $(F, \Delta)$-plane, showing additionally the curve $\text{HOM}_p$ (dark-blue) of Shilnikov bifurcation of the symmetric steady state $p$ with the Belyakov point $\textbf{BV}$, as well as selected curves (grey) of secondary Shilnikov bifurcations to $p$. (b1),(b2) Enlargements of regions near $\text{HOM}_p$ either side of the point $\textbf{BV}$, with dots indicating the locations of the phase portraits in Figs.~\ref{fig8} and~\ref{fig9}.}
\end{figure}

\subsection{\label{subsection5} Bifurcations of periodic orbits: period-doubling route to chaos}

Figure~\ref{fig3} shows the bifurcations of periodic orbits past the first Hopf bifurcation $\textbf{H}$, for $F=6.0$ and varying $\Delta$, with phase portraits in the $(P_1,P_2)$-plane showing a period-doubling route to chaos. Notice that panel~(a) is an enlargement of one of the branches of periodic orbits shown in Fig.~\ref{fig2}(e). The pair of periodic orbits that emerges at $\textbf{H}$ becomes unstable at the first period-doubling bifurcation $\textbf{PD}$, where a pair of stable periodic orbits with twice the period is created. As $\Delta$ is increased, the system goes through a first cascade of period-doubling bifurcations $\textbf{PD}$. Numerical integration shows that this cascade of period-doubling bifurcations $\textbf{PD}$ leads to the formation of a pair of chaotic attractors. Figure~\ref{fig3}(b1) depicts the phase portrait of the original pair of periodic orbits that emerge at the first Hopf bifurcation point $\mathbf{H}$, as indicated by the dashed line in panel~(a). The blue and orange dots represent the symmetric and asymmetric steady states $p$, and  $a_{\pm}$ created at the pitchfork bifurcation $\textbf{P}$. Due to the $\mathbb{Z}_2$-equivariance of system~\eqref{eq1}, the two periodic orbits near $a_+$ and $a_-$ are related by mirror symmetry through the diagonal $P_1=P_2$ of the $(P_1,P_2)$-plane of Fig.~\ref{fig3}(b). Panel~\ref{fig3}(b2) shows an enlargement of the periodic orbits near the steady state $a_+$. Past the first period-doubling bifurcation $\textbf{PD}$, we observe in Fig.~\ref{fig3}(c) a pair of periodic orbits with two loops as indicated by the dashed line in panel~(a). As $\Delta$ is increased further, the periodic orbits bifurcate infinitely many more times at period-doubling bifurcations $\textbf{PD}$ and a pair of chaotic attractors emerge, as shown in Fig.~\ref{fig3}(d) and indicated in panel~(a). After the point $(F,\Delta)=(6.0,6.5)$, the system goes through a reverse period-doubling cascade for increasing $\Delta$, where the number of loops of the stable pair of periodic orbits is halved at each point $\textbf{PD}$. Hence, the original pair of periodic orbits recovers its stability at about $\Delta=6.7$; see Fig.~\ref{fig3}(a).

Above $\Delta=6.8$, as in Fig.~\ref{fig3}(a), the original pair of periodic orbits undergoes two saddle-node bifurcations $\textbf{SP}$, leading to a $\Delta$-range with two coexisting pairs of stable periodic orbits. Moreover, one of the pairs of periodic orbits undergoes a second cascade of period-doubling bifurcations; hence, the points $\textbf{SP}$ delimit a range of multi-stability with coexistent periodic and chaotic attractors. Subsequently, the system undergoes further period-doublings until the branch of periodic orbits ends at the Shilnikov bifurcation point $\text{HOM}_p$, as shown in Fig.~\ref{fig3}(a).

Figure~\ref{fig4} shows chaotic attractors, a pair of Shilnikov bifurcations, and associated temporal traces for different values of $\Delta$ along the bifurcation diagram shown in Fig.~\ref{fig3}(a). In addition, we also show the positive branch of the one-dimensional \textit{unstable manifold} $W^u_+(p)$ of the saddle symmetric steady state $p$, which is the set of all points in phase space that converges to $p$ in backward time. The pair of chaotic attractors of Fig.~\ref{fig4}(a1) is the same shown in Fig.~\ref{fig3}(d). We note that $W_+^u(p)$ accumulates on the top-left chaotic attractor. Due to the  $\mathbb{Z}_2$ symmetry, the negative branch of the unstable manifold $W_-^u(p)$ of $p$ (not shown here) accumulates on the bottom-right chaotic attractor. As expected, the corresponding temporal traces of $P_{1}$ and $P_{2}$ of the pair of chaotic attractors in Fig.~\ref{fig4}(a2) show no repeating cycle. Notice that the attractors shown in Fig.~\ref{fig4}(a) are characterized by the dominance of one mode. Moreover, the amplitudes of oscillation of the output intensities $P_1$ and $P_2$ are out of phase. We refer to this type of dynamics as \emph{non-switching chaotic behavior}~\cite{andrus2021}.

Past the second period-doubling cascade as indicated by the dashed line $\Delta=6.9$ in Fig.~\ref{fig3}(a), we observe in Fig.~\ref{fig4}(b1) two larger chaotic attractors. Their temporal traces in Fig.~\ref{fig4}(b2) show more complex oscillations with epochs where the amplitudes of the output intensities $P_1$ and $P_2$ are equal. This is explained by the fact that the two attractors overlap near $p$ in the projection onto the $(P_1,P_2)$-plane of Fig.~\ref{fig4}(b1). However, they are still well separated in the full four-dimensional phase space by the stable manifold of $p$, since $W^u_+(p)$ still accumulates only on the top-left chaotic attractor. As $\Delta$ is increased to $\Delta=7.075$, as in  Fig.~\ref{fig4}(c1), the chaotic attractors change shape and show more overlap in the $(P_1,P_2)$-plane, yet still remain separated in phase space; this is reflected in the temporal traces in panel~(c2), which shows that the two attractors are very close to $p$ and, hence, the subspace $\text{Fix}_\eta$.

Indeed, just before the point $\text{HOM}_p$ in Fig.~\ref{fig3}(a), we find that, after its first loop around the chaotic attractor, the unstable manifold $W^u_+(p)$ makes several oscillations very close to $p$. At the point $\text{HOM}_p$, the unstable manifold $W^u(p)$ connects back to $p$ to create a pair of isolated homoclinic orbits of Shilnikov type. Figure~\ref{fig4}(d1) shows the Shilnikov homoclinic orbits in the $(P_1,P_2)$-plane, as computed with a numerical implementation of Lin's method~\cite{Krauskopf_2008}. The temporal profiles of $\text{HOM}_p$ in Fig.~\ref{fig4}(d2) depict two pulsed solutions, each localized in different regions of the phase space. Notice that the tail of the homoclinic orbit has small oscillations near $p$, which is a characteristic of Shilnikov bifurcation. Examination of the eigenvalue of $p$ at the point $\text{HOM}_p$ shows that this is a chaotic Shilnikov bifurcation, which is known to lead to chaotic dynamics~\cite{shill, shil2}. Thus, we expect the system to exhibit a plethora of exotic behavior near the point $\text{HOM}_p$. This includes sequences of saddle-node and period-doubling bifurcations.

\subsection{\label{subsection51} Global bifurcations}

We now focus on describing the organization of some key global bifurcations in the $(F,\Delta)$-plane. Figure~\ref{fig5} presents the two-parameter bifurcation diagram of system~\eqref{eq1} with additional curves of global bifurcations. The locus of the Shilnikov bifurcation from Fig.~\ref{fig2}(e) can be continued as a curve in the $(F,\Delta)$-plane. Figure~\ref{fig5}(a) shows the resulting curve $\text{HOM}_p$; it enters the shown region from the top, then has a fold of homoclinic bifurcation and exits from the right. Numerical continuation reveals a Belyakov transition point $\textbf{BV}$~\cite{bel,beltran2} on the curve $\text{HOM}_p$; this point is a codimension-two homoclinic bifurcation, where the homoclinic orbit changes from attracting to repelling. As we will discuss below in Sec.~\ref{belyakovsec}, there exist infinitely many more curves of homoclinic bifurcations near the curve $\text{HOM}_p$, of which some are shown in Figure~\ref{fig5}(a); see also the enlargements near Fig.~\ref{fig5} in panels~(b1) and~(b2). Our results are consistent with the unfolding of the point $\textbf{BV}$ as described by theory~\cite{bel}, but we find additional curves of homoclinic bifurcations on account of the $\mathbb{Z}_2$-equivariance of system~\eqref{eq1}. In this way, we clarify the relevance of the point $\textbf{BV}$ and these homoclinic bifurcations for the switching dynamics of periodic and chaotic solutions.

\section{\label{belyakovsec} Dynamics near the Belyakov transition}

\begin{figure}[H]
\centering
\includegraphics[]{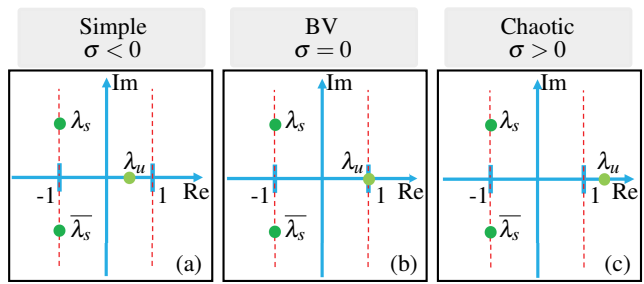}
\caption{\label{fig6} Illustration of different configurations of the leading eigenvalues of the symmetric steady states $p$ along the curve $\text{HOM}_p$. Shown here are the real unstable eigenvalue $\lambda_u$, and the two stable complex conjugates eigenvalues $\lambda_{s}$ and $\overline{\lambda}_{s}$.}
\end{figure}

As we have seen in Sec.~\ref{section2}, the bifurcation of the asymmetric steady states can lead to the formation of a unique pair of stable periodic orbits [Fig.~\ref{fig2}(d)], or to a $\Delta$-range where more than one pair of periodic orbits and chaotic attractors exist simultaneously [Fig.~\ref{fig2}(e)]. We find that the bifurcation responsible for the emergence or disappearance of chaotic dynamics is the Belyakov transition~\cite{bel,beltran2}, which occurs at the point $\textbf{BV}$.

\subsection{Transition from simple to chaotic dynamics} 

Near the curve $\text{HOM}_p$, the dynamics are strongly dependent on the eigenvalue configuration of $p$; more specifically, they depend on the relative strengths between the repelling and attracting eigendirections of $p$, also known as the \emph{saddle value}. As $p$ has three stable and one unstable eigenvalues, this leads to two generic configurations, one where the saddle $p$ is more strongly attracting than repelling; this configuration is referred to as ``simple." In the opposite case, where  the saddle $p$ is more strongly repelling than attracting, the configuration is referred to as ``chaotic." The codimension-two Belyakov transition $\textbf{BV}$ represents the moment where the Shilnikov bifurcation $\text{HOM}_p$ transitions from a simple to a chaotic Shilnikov bifurcation. 

The steady-state $p$ has three leading eigenvalues and one subleading eigenvalue.  The three leading eigenvalues are: an unstable real eigenvalue $\lambda_p$ and two stable complex-conjugate eigenvalues $\lambda_s$ and $\overline{\lambda_s}$. The subleading eigenvalue $\lambda_{ss}$ is stable and real. The saddle value $\sigma$ of $p$ is 
$$ \sigma = \text{Re}(\lambda_s) + \lambda_u.$$
The eigenvectors associated with the complex eigenvalues  $\lambda_s$ and $\overline{\lambda_s}$ span $\text{Fix}_\eta$; that is, these two eigenvalues govern the dynamics on this two-dimensional symmetry subspace. As the trace of the Jacobian of system~\eqref{eq1} is $-2$ when restricted to $\text{Fix}_\eta$, and $-4$ in the full system (see Appendix \ref{appendix1}),  the eigenvalues of $p$ satisfy
$$\text{Re}(\lambda_s)=-1, \text{ and }  \lambda_u +  \lambda_{ss} = -2 .$$
This implies that as $F$ and $\Delta$ vary, only the real eigenvalues $\lambda_u$ and $\lambda_{ss}$, and the imaginary part of $\lambda_s$ and $\overline{\lambda_s}$, change. Hence, the sign of the saddle value $\sigma=-1+ \lambda_u$, strictly depends on the eigenvalue $\lambda_u$, with the transition $\textbf{BV}$ occurring when $\lambda_u=1$.

The different configurations of leading eigenvalues of $p$ along the curve $\text{HOM}_p$ are shown in Fig.~\ref{fig6} . On the upper part of the curve $\text{HOM}_p$ we have $0< \lambda_u<1$ [Fig.~\ref{fig6}(a)], hence, the saddle value $\sigma$ is negative. This leads to a simple Shilnikov bifurcation with a unique (pair of) stable periodic orbit(s) for nearby parameter values. At the point $\textbf{BV}$, $\lambda_u=1$ [Fig.~\ref{fig6}(b)], the system exhibits the Belyakov transition. Past the point $\textbf{BV}$ for increasing $F$, we have $\lambda_u>1$ as in Fig.~\ref{fig6}(c), and the saddle value $\sigma$ is positive. Hence, the Shilnikov bifurcation is chaotic~\cite{shill,shil2} and there are infinitely many periodic orbits at nearby parameter values. In the case of system~\eqref{eq1}, these periodic orbits bifurcate through period-doubling cascades to create different chaotic attractors, as was shown in Figs.~\ref{fig4} and ~\ref{fig5}. 

\subsection{\label{simple_Shil_sec} Dynamics near the simple Shilnikov bifurcation}

\begin{figure}
\centering
\includegraphics[]{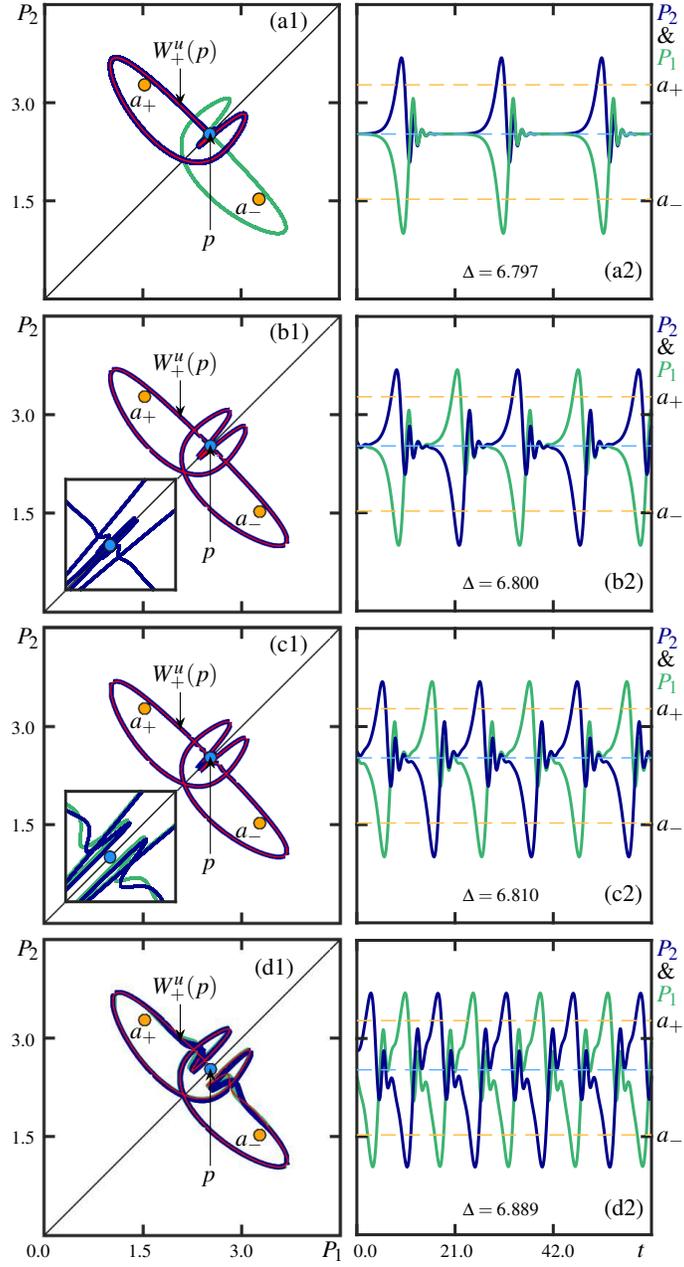}
\caption{\label{fig8} (a1)--(d1) Pairs of periodic and chaotic attractors, and (a2)--(d2) their associated temporal traces for $F=4.0$ and $\Delta=6.797$, $\Delta=6.800$, $\Delta=6.810$, $\Delta=6.889$, respectively. Also shown in (a1)--(d1) is the positive branch $W^+_u ( p)$ (red curves) of the unstable manifold of $p$.}
\end{figure}

To understand the relevance of the simple Shilnikov bifurcation $\text{HOM}_p$ for the periodic orbits, we consider a slice of constant $F$ through the two-parameter bifurcation diagram to the left of the point $\textbf{BV}$ and close to the curve $\text{HOM}_p$. Figure~\ref{fig5}(b1) shows an enlargement of the $(F,\Delta)$-plane close to the curve $\text{HOM}_p$, with a dashed line of fixed pump power at $F=4.0$. Figure~\ref{fig8} shows the phase portraits and temporal traces of the output intensities $P_1$ and $P_2$ at the points labeled 7(a) to 7(d) in Fig.~\ref{fig5}(b1).

Panel~\ref{fig8}(a1) shows (a pair of) stable periodic orbits very close to the curve $\text{HOM}_p$. Here, after each loop of oscillations at the top-left of the $(P_1,P_2)$-plane, the positive branch of the unstable manifold $W^u_+(p)$ comes very close the steady state $p$, and likewise for the negative branch of the unstable manifold  $W^u_-(p)$ (not shown here). Hence, the pair of periodic orbits are nearly homoclinic. This means that each oscillation spends a long time near $p$, leading to two persistent temporal trains of pulse-like solutions of large periods [Fig.~\ref{fig8}(a2)]. We note that the two attractors shown in Fig.~\ref{fig8}(a1) are well separated by the (three-dimensional) stable manifold of $p$ in the full four-dimensional phase space.

Further away from the curve $\text{HOM}_p$, we observe that the positive branch of the unstable manifold $W^u_+(p)$ crosses the symmetry subspace $\text{Fix}_{\eta}$ to accumulate, on an S-invariant periodic orbit [Fig.~\ref{fig8}(b1)] that ``connects" the two regions of the phase space previously separated. Notice that this periodic orbit is close in shape to the union of the two periodic orbits in Fig.~\ref{fig8}(a1). Therefore, the simple Shilnikov bifurcation $\text{HOM}_p$ acts as a gluing bifurcation~\cite{glue1} that connects the two periodic orbits separated by the stable manifold of $p$. Figure~\ref{fig8}(b2) shows the temporal traces of $P_1$ and $P_2$ of the attractor in panel~\ref{fig8}(b1), which display regular switching between the two regions of the phase space containing the steady states $a_+$ and $a_-$, respectively. 

As $\Delta$ is increased, and we move further from the homoclinic bifurcation curve, the period of oscillation decreases. Moreover, the symmetric periodic orbit created at $\text{HOM}_p$ undergoes a pitchfork bifurcation of periodic orbits, where two asymmetric, stable periodic orbits are created; hence, the pitchfork bifurcation acts as a symmetry-breaking bifurcation of periodic orbits. The pair of asymmetric periodic orbits is shown in Fig.~\ref{fig8}(c1), but it is difficult to distinguish them. This is because they exist very close to the symmetry-breaking bifurcation in the parameter space. The insets of panels~(b1) and~(c1) show enlargements near the symmetric steady state $p$ to clarify the difference and to distinguish the two asymmetric periodic orbits. Their temporal traces in Fig.~\ref{fig8}(c2) show that the output intensities of $P_1$ and $P_2$ still exhibit regular switching behavior. Moreover, we note that each of the output intensities $P_1$ and $P_2$ oscillates with a slightly larger amplitude in the regions containing $a_+$ and $a_-$, respectively. As $\Delta$ is increased further, the two periodic orbits undergo a period-doubling cascade; this fact also confirms that the system went through a symmetry-breaking bifurcation of periodic orbits~\cite{con1,con2}. This period-doubling cascade leads to the formation of two asymmetric chaotic attractors, as in Fig.~\ref{fig8}(d1). Their temporal traces are shown in Fig.~\ref{fig8}(d2), and they still show regular switching; the maxima of the respective output intensity, on the other hand, are actually irregular (due to chaos on a small scale that is not resolved on the scale of the shown time trace).

\subsection{Infinitely many bifurcations near the chaotic Shilnikov bifurcation} 

Loosely speaking, theory predicts the existence of unaccountably many families of homoclinic orbits near the chaotic Shilnikov bifurcation~\cite{bel,shill}. Therefore, we expect an accumulation of infinitely many Shilnikov bifurcations for nearby parameter values. To investigate the dynamics of different families of homoclinic bifurcations, we use a numerical implementation of Lin's method~\cite{Krauskopf_2008} to identify them and to find their curves of existence in the $(F,\Delta)$-plane. 

\begin{figure}
\centering
\includegraphics{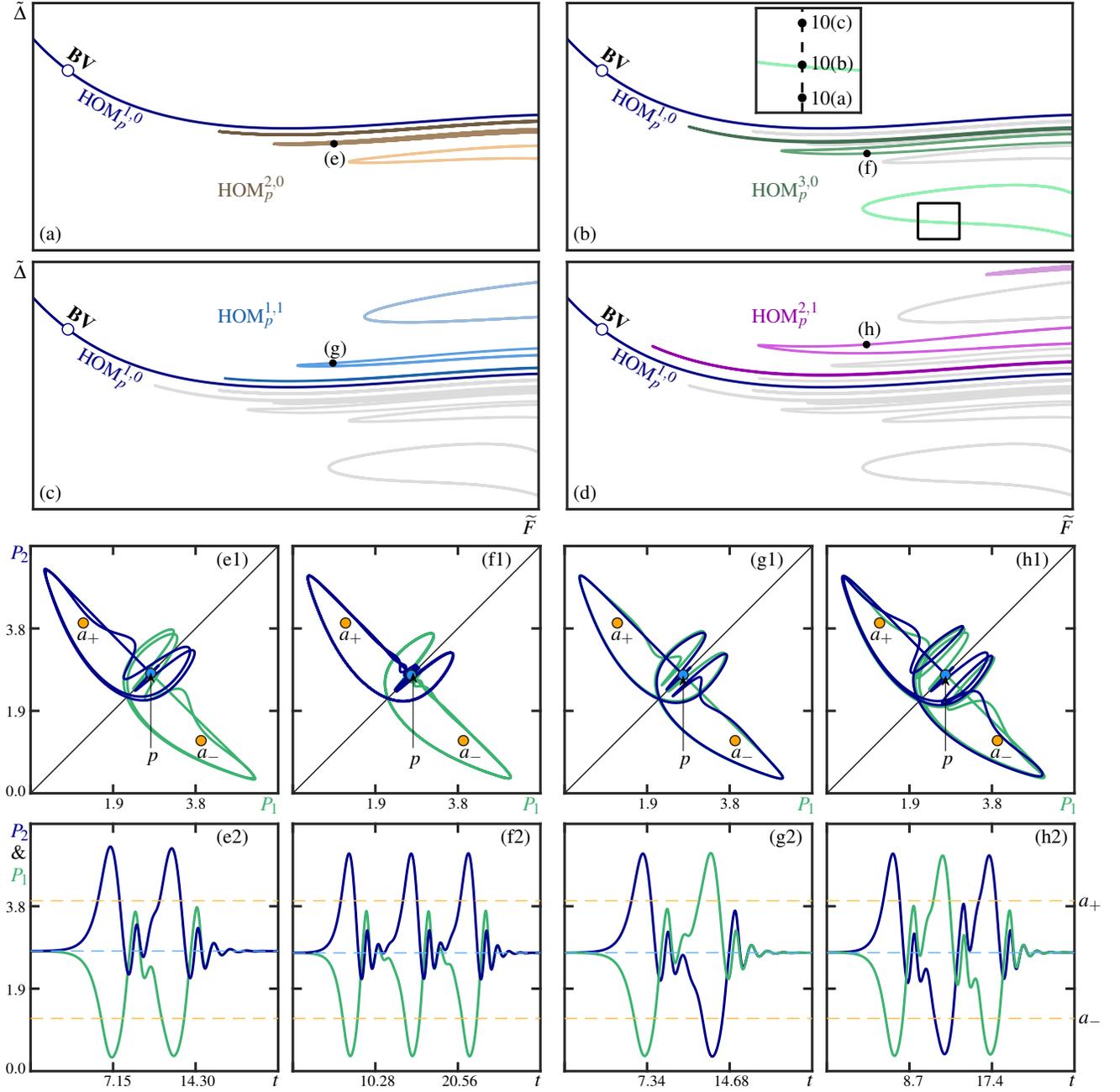}
\caption{\label{fig7} Unfolding of (a)--(b) non-switching  and (c)--(d) switching Shilnikov homoclinic orbits of system~\eqref{eq1}, near $\textbf{BV}$. Shown are the curves of $(2,0)$-homoclinic $\text{HOM}^{2,0}_{p}$ (brown), $(1,1)$-homoclinic  $\text{HOM}^{1,1}_{p}$ (blue), $(3,0)$-homoclinic $\text{HOM}^{3,0}_{p}$ (green) and $(2,1)$-homoclinic  $\text{HOM}^{2,1}_{p}$ (purple) bifurcations. (e1)--(h1) Pair of Shilnikov $(m,n)$-homoclinic orbits and (e2)--(h2) corresponding temporal profiles for $F=6.25$ and $\Delta=7.1779$, $\Delta=7.0792$, $\Delta=7.0921$ and $\Delta=7.1106$, respectively.}
\end{figure}

We consider here homoclinic orbits for which the unstable manifold $W^u_+(p)$ makes $m$ loops around $a_+$ and $n$ loops around $a_-$ before returning to $p$. For notational convenience, we now refer to the associated bifurcation curves with reference to $m$ and $n$. For example, the homoclinic orbit along $\text{HOM}_{p}$ will be referred to as a $(1,0)$-homoclinic orbit, and this is now denoted by $\text{HOM}^{1,0}_{p}$. As we will see, there are multiple curves corresponding to the same pair $(m,n)$. The panels of Fig.~\ref{fig7} show different sets of computed curves of homoclinic bifurcations, associated phase portraits, and temporal profiles near the codimension-two point $\textbf{BV}$. For better visualization, we map the $(F, \Delta)$-plane to the $(\tilde{F},\tilde{\Delta})$-plane, by rotation over the angle $-\pi/7$ around the point $\textbf{BV}$.

\subsubsection{Non-switching homoclinic bifurcations \label{sec5a}}

We refer to homoclinic orbits with $n=0$ as \textit{non-switching homoclinic orbits}. System~\eqref{eq1} exhibits many non-switching $(2,0)$-Shilnikov bifurcations $\text{HOM}^{2,0}_{p}$, whose loci can be continued as curves in the $(F,\Delta)$-plane. Figure~\ref{fig7}(a) shows three such curves out of the infinitely many that exist near $\text{HOM}^{1,0}_{p}$; each curve $\text{HOM}^{2,0}_{p}$ enters the shown region from the right, then has a fold of homoclinic bifurcation with respect to $\tilde{F}$ and turns back to the right. As predicted by theory~\cite{bel}, the two branches of these homoclinic bifurcation curves are closer together as $\text{HOM}^{1,0}_{p}$ is approached. Moreover, the fold points of the curves $\text{HOM}^{2,0}_{p}$ accumulate on the codimension-two point $\textbf{BV}$. Figure~\ref{fig7}(e1) shows the phase portraits (of a pair) of the non-switching $(2,0)$-homoclinic bifurcations at the point (e) indicated in panel~\ref{fig7}(a). Notice that each homoclinic orbit makes two loops and then connects to the steady state $p$ in the symmetry subspace $\text{Fix}_{\eta}$. The temporal profiles in Fig.~\ref{fig7}(e2) show pulsed solutions with two peaks corresponding to the number of loops.

 Figure~\ref{fig7}(b) shows the curves of non-switching $(3,0)$-homoclinic bifurcations $\text{HOM}^{3,0}_{p}$ together with the curves $\text{HOM}^{2,0}_{p}$ in the background. We observe that the curves $\text{HOM}^{3,0}_{p}$ are located in between the curves $\text{HOM}^{2,0}_{p}$ as predicted by the theoretical unfolding ~\cite{bel, beltran1}. We show in Fig.~\ref{fig7}(f1) the phase portraits of a pair of non-switching $(3,0)$-homoclinic orbits at the point (f) in Fig.~\ref{fig7}(b). Since the point (f) is very close to the curve $\text{HOM}^{1,0}_{p}$, one can hardly distinguish the three loops of each homoclinic orbit. However, the temporal profiles in Fig.~\ref{fig7}(f2) clearly show pulsed solutions with three prominent peaks.

\subsubsection{Switching homoclinic bifurcations}

Previous studies~\cite{bel,beltran2} of the codimension-two point $\textbf{BV}$ concern systems without $\mathbb{Z}_2$-symmetry. In this case, it is expected that the curves of homoclinic bifurcations in the unfolding should accumulate on one side of the primary homoclinic curve $\text{HOM}^{1,0}_{p}$. This is indeed the case for the bifurcation curves of the non-switching homoclinic orbits. However, due to $\mathbb{Z}_2$-equivariance, we find infinitely many curves accumulating onto the primary homoclinic curve $\text{HOM}_p$ also from the other side. In particular, these bifurcations generate homoclinic orbits~\cite{shila,glen} that alternate between the two regions of the phase space separated by the stable manifold of $p$, that is, homoclinic orbits with both $m$ and $n$ nonzero. Therefore, we refer to these homoclinic bifurcations as \emph{switching homoclinic bifurcations}. 

Figure~\ref{fig7}(c) shows curves of $(1,1)$-homoclinic bifurcations $\text{HOM}^{1,1}_{p}$ together with the curves of non-switching homoclinic bifurcations in the background. Notice that the curves $\text{HOM}^{1,1}_{p}$  are located above the curve $\text{HOM}^{1,0}_{p}$. They also accumulate on this primary curve of homoclinic bifurcation, and their organization is very similar to the curves of non-switching $(2,0)$-homoclinic bifurcations. However, in the phase space [Fig.~\ref{fig7}(g1)], each of the switching $(1,1)$-homoclinic orbits has one loop around each of the asymmetric steady states $a_+$ and $a_-$, and then connect back to the steady state $p$ in the symmetry subspace $\text{Fix}_{\eta}$. In the temporal profiles [Fig.~\ref{fig7}(g2)], these loops translate to large jumps in the amplitude of the output intensities $P_1$ and $P_2$ from the region containing $a_+$ to the region with $a_-$, and vice versa. 

Additionally, we show in Fig.~\ref{fig7}(d) the curves of switching $(2,1)$-homoclinic bifurcations $\text{HOM}^{2,1}_{p}$, again together with the curves presented in panels~(a)--(c). We notice that the curves $\text{HOM}^{2,1}_{p}$ are located between the curves $\text{HOM}^{1,1}_{p}$, and are organized as the curves $\text{HOM}^{3,0}_{p}$. Figures~\ref{fig7}(h1) and~(h2) show the phase portraits and temporal profiles of $(2,1)$-homoclinic orbits at the point (h) indicated in Fig.~\ref{fig7}(d). Notice that each homoclinic orbit makes three oscillations around the steady states $a_+$ and $a_-$. 

\subsection{Dynamics near $\textbf{BV}$: Infinitely many homoclinic curves and periodic solutions}

Switching homoclinic bifurcations in $\mathbb{Z}_2$-equivariant dynamical systems have been extensively studied both theoretically~\cite{glen,shila} and numerically~\cite{andrus2021,barrio} for the case of systems with chaotic Shilnikov bifurcations. However, the unfolding of homoclinic bifurcations near a $\mathbb{Z}_2$-equivariant Belyakov transition has yet to be studied to the best of our knowledge. The bifurcation diagram in Fig.~\ref{fig7}(a)--(d) presents a partial numerical unfolding of homoclinic bifurcations near a $\mathbb{Z}_2$-equivariant codimension-two Belyakov homoclinic bifurcation. 

We note that each homoclinic orbit involved in the unfolding of the codimension-two point $\textbf{BV}$ has a negative saddle value. Hence, they are organizing centers of further families of subsidiary homoclinic bifurcation curves. Therefore, the complete picture of the homoclinic curves near the point $\textbf{BV}$ is even more complicated. Moreover, each homoclinic orbit in the unfolding of the point $\textbf{BV}$ is associated with a family of stable or unstable periodic solutions. Indeed, a point on each homoclinic bifurcation curve is associated with a branch of periodic orbits when one parameter is varied, as in Fig.~\ref{fig3}(a). We observe and conjecture that infinitely many saddle periodic and chaotic attractors exist near the point $\textbf{BV}$.

\subsection{Dynamics near the chaotic Shilnikov bifurcation $\text{HOM}^{1,0}_p$}

We now turn our attention to the relevance of the Shilnikov bifurcation $\text{HOM}^{1,0}_p$ for the dynamics of the output intensities $P_1$ and $P_2$ to the right of the point $\textbf{BV}$. Below the curve $\text{HOM}^{1,0}_p$, the two branches of the unstable manifold $W^u_+(p)$ and $W^u_-(p)$ accumulate, respectively, on two different chaotic attractors [Fig.~\ref{fig4}(c1)] located in the region of the phase space containing the asymmetric steady states $a_+$ and $a_-$, as discussed in Sec.~\ref{subsection5}. However, above the curve $\text{HOM}^{1,0}_p$, we find that each branch of the unstable manifold $W^u(p)$ visits both regions of the space containing the asymmetric steady states $a_+$ and $a_-$, respectively. 

\begin{figure}[H]
\centering
\includegraphics[]{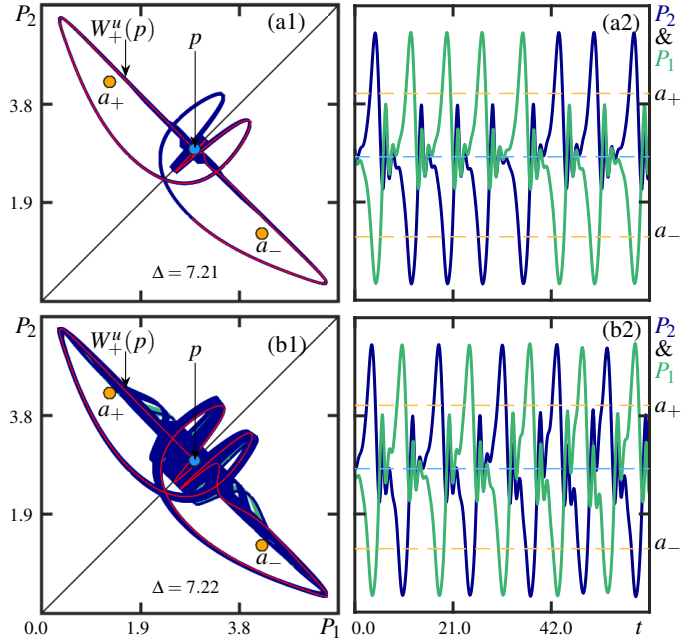}
\caption{\label{fig9} (a1)--(b1) Pairs of switching chaotic attractors and (a2)--(b2) their associated temporal traces for $F=6.26$ and $\Delta=7.21$, $\Delta=7.229$, respectively. Also shown in (a1)--(b1) is the positive branch $W^+_u ( p)$ of the unstable manifold of $p$.} 
\end{figure}

Figure~\ref{fig9} shows phase portraits and temporal traces of the output intensities $P_1$ and $P_2$, and the positive branch of  the unstable manifold $W^u_+(p)$ above the curve $\text{HOM}^{1,0}_p$, for fixed $F$ and the two values of $\Delta$ indicated in Fig.~\ref{fig5}(b2). For $\Delta=7.21$ as in Figure~\ref{fig9}(a1), the positive branch of the unstable manifold $W^u_+(p)$ does a first loop around $a_+$, then a second one around $a_-$, before oscillating irregularly. More precisely, after its first two loops, $W^u_+(p)$ oscillates for a considerable time in the region containing $a_-$, before switching to the region containing $a_+$, where it also performs many oscillations. Over longer timescales, we observe that $W^u_+(p)$ transitions irregularly between these two regions. Hence, we observe the formation of a large symmetric chaotic attractor that switches between the two regions of the phase space containing the steady states $a_+$ and $a_-$. Figure~\ref{fig9}(a2) shows the temporal traces of $P_2$ and $P_1$ associated to the chaotic attractor in panel~\ref{fig9}(a2). As expected, the amplitudes of the output intensities of $P_2$ and $P_1$  exhibit a chaotic intermittent switching between the two asymmetric steady states $a_+$ and $a_-$. As we will see later, this transition from non-switching to switching chaotic behavior can also be found near other wild Shilnikov bifurcations. This process is referred to in the literature as symmetry increasing of chaotic attractors~\cite{CHOSSAT,PAshwin}.

For $\Delta=7.22$ as in Fig.~\ref{fig9}(b1), we observe the formation of two large asymmetric chaotic attractors. However, the temporal profiles [Fig.~\ref{fig9}(b2)] associated with these attractors show regular switching between both regions of the phase space containing $a_+$ and $a_-$, respectively. We now observe two attractors showing that the system underwent a symmetry-breaking bifurcation. Indeed, as the detuning is increased from $\Delta =7.21$, the system first displays a symmetric pair of periodic orbits [similar to Fig.~\ref{fig8}(b)] that undergoes a pitchfork bifurcation where two asymmetric periodic orbits are created. This pitchfork bifurcation can be continued as a curve in the $(F,\Delta)$-plane and connects back to the previous pitchfork bifurcation mentioned in Sec.~\ref{simple_Shil_sec}, for the case where the curve $\text{HOM}^{1,0}_p$ is associated with simple dynamics. After the pitchfork bifurcation, the newly created pair of asymmetric periodic orbits undergoes infinitely many global bifurcations, leading to the formation of a pair of chaotic attractors with more regular switching as in Fig.~\ref{fig9}(b1). We note that the chaotic attractors in Fig.~\ref{fig9}(b) emerge from a cascade of period-doubling bifurcations of periodic orbits associated with one of the $(1,1)$-homoclinic curve in Fig.\ref{fig7}(c). Interestingly, their temporal traces have a similar switching pattern. This is further evidence that homoclinic bifurcations organize the different switching patterns observed in system~\eqref{eq1}.

\subsection{Dynamics near the homoclinic bifurcations $\text{HOM}^{m,0}_p$}

\begin{figure}
\centering
\includegraphics[]{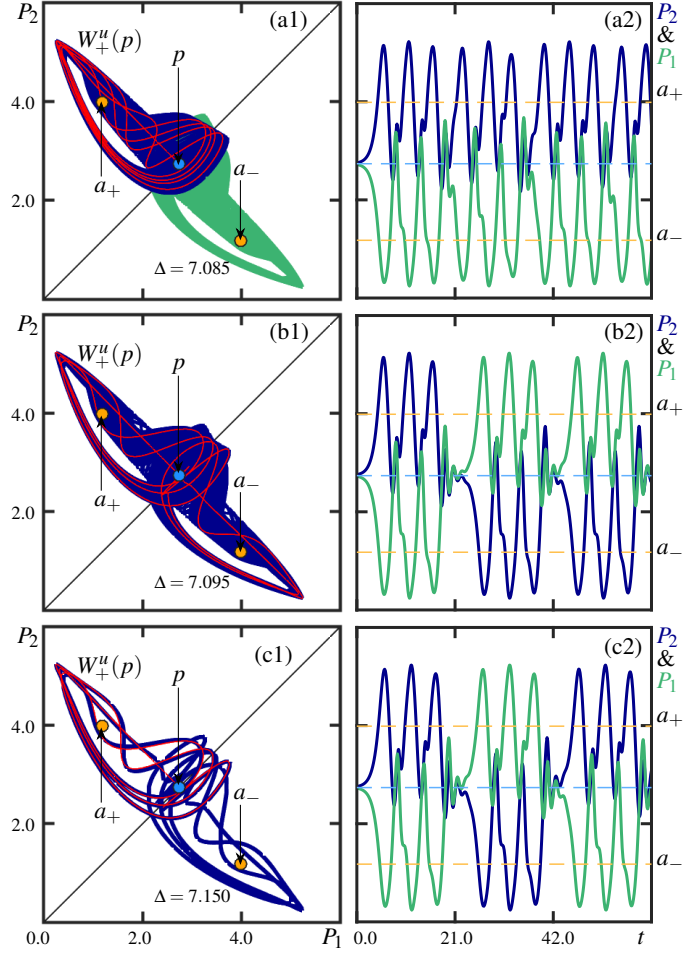}
\caption{\label{fig10} (a1)--(c1) Phase portraits and (a2)--(c2) associated temporal traces for $F=6.200$ and $\Delta=7.085$, $\Delta=7.095$, $\Delta=7.150$, respectively. Also shown in (a1)--(b1) is the positive branch $W^+_u ( p)$ of the unstable manifold of $p$.} 
\end{figure}

It is now clear that the homoclinic bifurcations of the locus  $\text{HOM}^{1,0}_p$ play a key role in the formation of different dynamics in system~\eqref{eq1}. Hence, we are now intrigued by the relevance of each curve in $\text{HOM}^{m,0}_p$. Figure~\ref{fig10} shows the phase portraits and temporal traces of $P_1$ and $P_2$ for fixed $F$, and different values of $\Delta$ near the lower curve $\text{HOM}^{3,0}_p$ in Fig.~\ref{fig7}(b), as indicated by the inset. For $\Delta=7.085$ as in Fig.~\ref{fig10}(a1), we observe that each branch of the unstable manifold of $p$ accumulates on two different non-switching chaotic attractors, which overlap near $p$ in this projection. Their temporal traces in Fig.~\ref{fig10}(a2) show that the mode $P_2$ is always dominant, but there are epochs of time where both amplitudes are equal. By looking carefully, we also note that the oscillations of $P_1$ and $P_2$ show a hint of period-three dynamics. This is explained by the fact that the pair of chaotic attractors shown in Fig.~\ref{fig10}(a) is obtained from a period-doubling cascade of a pair of non-switching periodic orbits with three loops.

For $\Delta=7.100$ as in Fig.~\ref{fig10}(b1), the positive branch $W^+_u ( p)$ of the unstable manifold of $p$ accumulates on a large and symmetric chaotic attractor that switches between the two regions of the phase space containing the steady states $a_+$ and $a_-$. Notice that this chaotic attractor is close in shape to the union of the two chaotic attractors in Fig.~\ref{fig10}(a1). Figure~\ref{fig10}(b2) shows the temporal traces of the attractors in panel~\ref{fig10}(b1); as expected, $P_1$ and $P_2$ display intermittent switching behaviors between the two regions. Notice that $P_2$ has three oscillations in the region of the phase space containing $a_+$, before switching to the region containing $a_-$, and likewise for $P_1$. This is characteristic of the Shilnikov bifurcation $\text{HOM}^{3,0}_p$, as is evidenced by the period-three nature of the dynamics. Therefore, the Shilnikov bifurcation $\text{HOM}^{3,0}_p$ also involves symmetry increasing of chaotic attractors. Our results show that this phenomenon in system~(\ref{eq1}) involves crossing infinitely many homoclinic bifurcation curves that accumulate on a central curve chaotic Shilnikov bifurcation; namely, $\text{HOM}_p^{1,0}$, in Fig.~\ref{fig9},  and $\text{HOM}_p^{3,0}$, in Fig.\ref{fig10}. We speculate that this transition is rather complicated and involves tangency bifurcations between global invariant manifolds of $p$ and those of different saddle periodic orbits ---akin to similar transitions studied in Ref.~\cite{andrus2021}. 

As $\Delta$ increases, the system undergoes many global bifurcations, and we observe the formation of a large symmetric periodic orbit about $\Delta=7.150$; see Fig.~\ref{fig10}(c1). This indeed confirms that the two chaotic attractors that merged at the Shilnikov bifurcation $\text{HOM}^{3,0}_p$, are obtained through period-doubling cascades of a pair of non-switching periodic orbits with three loops. The corresponding temporal traces of $P_1$ and $P_2$ in Fig.~\ref{fig10}(c2) display intermittent oscillations with periodic switching between the two regions of the phase space containing $a_+$ and $a_-$, every three cycles.

In general, we observe that each homoclinic bifurcation $\text{HOM}^{m,0}_p$ of system~\eqref{eq1}, represents the moment where two non-switching chaotic attractors merge to form a switching chaotic attractor or a periodic orbit with regular switching every $m$ cycles. Additionally, Fig.~\ref{fig10} shows that each curve of homoclinic bifurcation $\text{HOM}^{m,0}_p$ enclosed a region of the $(F,\Delta)$-plane, where one can find infinitely many switching homoclinic bifurcations $\text{HOM}^{m,n}_p$, as was illustrated with the curve $\text{HOM}^{1,0}_p$.

\section{\label{conclusion} Discussion and conclusions}

We have investigated the interaction of two optical fields in a single-mode ring resonator, emphasizing on regular and chaotic self-switching oscillations, as described by the $\mathbb{Z}_2$-equivariant vector field~(\ref{eq1}). We first derived analytical expressions for all steady-state bifurcations in terms of all three system parameters: intensity $F$ and detuning $\Delta$ of the pulses, and the ratio $B$ of their cross- versus self-phase modulation. These results agree with and extend the expressions for bistability and spontaneous symmetry breaking of symmetric steady states in previous studies~\cite{woodley2018,lewis2020}, and they allow us to plot the associated bifurcation curves in the $(F,\Delta)$-plane for any value of $B$ directly. 

We then studied bifurcations of non-switching periodic orbits that emerge from Hopf bifurcations and found that they may undergo a period-doubling cascade to create non-switching chaotic attractors. Both of these objects, which come in pairs, can then undergo Shilnikov bifurcations of saddle symmetric steady states $p$ to regular or chaotic self-switching dynamics. More specifically, on the curve $\text{HOM}_p$ of the main Shilnikov bifurcation, we found a Belyakov point $\textbf{BV}$, where the saddle quantity of $p$ is zero. The point $\textbf{BV}$ is an organizing center for infinitely many Shilnikov bifurcation curves nearby. Our computed unfolding in the $(F,\Delta)$-plane near this point $\textbf{BV}$ is in agreement with the theory~\cite{beltran1,beltran2} for the generic case but reveals an even richer bifurcation diagram with additional sets of Shilnikov bifurcations due to the $\mathbb{Z}_2$-equivariance. 

Transition through each of these Shilnikov bifurcations increases the symmetry of the attractor, from either two periodic solutions to a single periodic solution, or from two chaotic attractors to a single chaotic attractor. In particular, these bifurcations mark the transition from non-switching dynamics, with the (average) power concentrated in one of the two electric fields, to self-switching, which features regular episodes of irregular switching between one or the other field having the higher power. To be more specific, the main transition to intermittent switching~\cite{selfswi} is organized by the Belyakov transition point $\textbf{BV}$ on the curve $\text{HOM}_p$. We observed that the ensuing large chaotic attractor undergoes infinitely many global bifurcations, resulting in the emergence of stable switching periodic orbits of large periods corresponding to different switching patterns. It should be possible to observe experimentally some of the switching dynamics described in this paper, especially near the point $\textbf{BV}$, where infinitely many types of switching behaviors exist in a sufficiently large parameter range. Indeed, in Ref.~\cite{selfswi} the authors observed experimentally that the system can exhibit near-periodic and chaotic intermittent switching behaviors for specific parameter values. Moreover, in regions of the $(F,\Delta)$-plane where $p$ is more attracting than repelling, we expect the pulse-liked periodic non-switching and switching oscillations to be within experimental reach, provided the system is symmetrically balanced. Our overall bifurcation diagram in the $(F,\Delta)$-plane provides a road map that may help guide experimental studies to find such exotic behavior in this optical system.


Our findings suggest a number of research questions that will be addressed in future work. First of all, it will be interesting to focus on the sensitivity of chaotic switching oscillations by mapping out further their regions of existence as well as regions of stability. Secondly, preliminary results show that system~(\ref{eq1}) exhibits more exotic dynamical behaviors organized by a codimension-two point concerning the existence of a heteroclinic cycle, which will be reported elsewhere. As for the Belyakov point studied here, these results agree with theoretical work for the generic case~\cite{impo1,shas}, but we also find additional global bifurcations because of $\mathbb{Z}_2$-equivariance.


\begin{acknowledgments}
A. Giraldo was supported by KIAS Individual Grant No.CG086101, at Korea Institute for Advanced Study. The authors thank Pascal Del'Haye and Gian-Luca Oppo for valuable discussions, including on the experimental realization. We also thank Kevin Stitely for helpful input and insightful comments.
\end{acknowledgments}

\appendix

\section{\label{appendix1} Jacobians in Cartesian coordinates}


System~\eqref{eq1} can be written as the four-dimensional real vector field
\begin{equation}
\begin{aligned}
\frac{d X_1}{d t} &= \sqrt{F} - X_1 + (X_1^2 + Y_1^2 + B X_2^2 + B Y_2^2 - \Delta) Y_1,
\\
\frac{d X_2}{d t} &= \sqrt{F} - X_2 + (X_2^2 + Y_2^2 + B X_1^2 + B Y_1^2 - \Delta)Y_2,
\\
\frac{d Y_1}{d t} &= - Y_1 - (X_1^2 + Y_1^2 + B X_2^2 + B Y_2^2 - \Delta)X_1,
\\
\frac{d Y_2}{d t} &= - Y_2 - (X_2^2 + Y_2^2 + B X_1^2 + B Y_1^2 - \Delta)X_2.
\label{append1}
\end{aligned}
\end{equation}
where $E_{1,2} = X_{1,2} + i Y_{1,2}$. Its Jacobian is
\begin{equation}
  \label{append2} 
  \begin{aligned}
   J &= 
   \begin{pmatrix}
    2  X_1 Y_1-1  & C_1 & 2 B X_2 Y_1 & 2 B Y_2 Y_1 \\
    D_1 & - 2  X_1 Y_1 -1  & -2 B X_2 X_1 & -2 B Y_2 X_1 \\
    2 B X_1 Y_2 & 2 B Y_1 Y_2 & 2  X_2 Y_2 -1  & C_2 \\
    -2 B X_1 X_2 & -2 B Y_1 X_2 & D_2 & - 2  Y_2 X_2 -1 
   \end{pmatrix}.
  \end{aligned}
\end{equation}
where $C_{1,2} (X_{1,2}, Y_{1,2}) = 3  Y_{1,2}^2 +  X_{1,2}^2 + B X_{2,1}^2 + B Y_{2,1}^2 - \Delta$, and $D_{1,2}= -3  X_{1,2}^2 -  Y_{1,2}^2 - B X_{2,1}^2 - B Y_{2,1}^2 + \Delta$.

From the expression of $J$, we obtain that the divergence of system~\eqref{append1} is always equal to $-4$.


In the fixed-point subspace $\text{Fix}_\eta$ we have $X=X_1=X_2$ and $Y=Y_1=Y_2$, and the reduced system~\eqref{eqt101} becomes
\begin{equation}
\begin{aligned}
\frac{d X}{d t} &= \sqrt{F} - X + [(B + 1)X^2 + (B+1)Y^2 - \Delta] Y,
\\
\frac{d Y}{d t} &= - Y - [(B + 1)X^2 + (B+1)Y^2 - \Delta]X, 
\label{append3}
\end{aligned}
\end{equation}
with Jacobian 
\begin{equation}
  \label{eqt11} 
  \begin{aligned}
   J_S= 
   \begin{pmatrix}
    -1 + 2(B+1)XY & (B+1)(X^2 + 3 Y^2) - \Delta \\
    -(B+1)(3X^2 + Y^2) + \Delta & -1 - 2(B+1) XY
   \end{pmatrix}
  \end{aligned}
\end{equation}
The eigenvalues of $J_S$ are
$$\lambda_s=-1 \pm i \sqrt{a},$$
with 
$$
a=\Delta^2 + 4\Delta(B+1)(X^2+Y^2)+3(B+1)^2(X^4+Y^4+2X^2Y^2).
$$
In particular, the trace of $J_S$ is $-2$; hence, the divergence of system~\eqref{append3} is always negative. Therefore, there cannot be periodic solutions in the two-dimensional invariant subspace $\text{Fix}_\eta$, which means that any symmetric periodic orbits of system~\eqref{eq1} are $S$-symmetric.

\section{\label{appendix2} Stability analysis of asymmetric steady states}

The steady states of system~\eqref{eq1} can be obtained by solving
\begin{equation}
\begin{aligned}
\sqrt{F} &= [ 1 - i( |E_{1}|^2+B |E_{2}|^2 -\Delta)] E_{1},
\\
\sqrt{F} &= [ 1 - i( |E_{2}|^2+B |E_{1}|^2 -\Delta)] E_{2}.
\end{aligned}
\label{apb1}
\end{equation} 
By multiplying each equation of the coupled system \eqref{apb1} with its complex conjugate, we obtain
\begin{subequations}
\begin{align}
F &= P_{1} + (P_{1} + B P_{2} -\Delta)^2P_{1},
\label{apb21}\\
F &= P_{2} + (P_{2} + B P_{1} -\Delta)^2P_{2},
\label{apb22}
\end{align}
\label{apb2}
\end{subequations} 
where $P_{1,2}=|E_{1,2}|^2$. 
%
%
Adding Eqs.~\eqref{apb21} and~\eqref{apb22} gives
\begin{equation}
F=\frac{[(B+1)(P_1 + P_2) -2\Delta][(P_1 + P_2 - \Delta)^2 + 1]}{B - 1},
\label{apb3}
\end{equation}
which yields Eq.~\eqref{eq:13} with the transformation $S=P_1+P_2$.


By equating Eqs.~\eqref{apb21} and~\eqref{apb22}, one obtains
\begin{equation}
0=(P_1 - P_2)[(P_1 + P_2 - \Delta)^2 + 1 - (B-1)^2 P_1 P_2].
\label{apb4}
\end{equation}
Given that we are interested in asymmetric solutions, we consider $P_1\neq P_2$ and rewrite Eq.~\eqref{apb4} as
\begin{equation}
0= (S - \Delta)^2 + 1 - (B-1)^2 P_1 P_2.
\label{apb5}
\end{equation}
Note that, by solving Eqs.~\eqref{eq:13} and~\eqref{apb5} simultaneously, one can obtain explicit (but lengthy) expressions for the output powers $P_{1}$ and $P_{2}$ of the asymmetric steady states.


\bibliography{DBGBK}

\end{document}